\documentclass[review]{elsarticle}

\usepackage{lineno,hyperref}
\modulolinenumbers[5]
\usepackage{placeins}
\usepackage{amsmath,amsthm,amssymb,epsfig,bm,amsmath}
 \usepackage{graphicx}
 \usepackage{subcaption}
\usepackage{float}
\usepackage{csquotes}
\usepackage{amsmath}
\usepackage{mhchem}
\usepackage{xparse}
\usepackage{adjustbox}
\usepackage{MnSymbol}
\usepackage{mathdots}
\usepackage{makecell}
\usepackage{listings}
\usepackage{footnote}
\usepackage[flushleft]{threeparttable}
\usepackage{multirow}
\usepackage{relsize}
\usepackage{lettrine}
\usepackage{courier}
\usepackage{color} 
\definecolor{mygreen}{RGB}{28,172,0} 
\definecolor{mylilas}{RGB}{170,55,241}
\usepackage{amsmath,amsthm,amssymb,amsfonts} 
\usepackage{graphicx} 
\usepackage[margin=1in]{geometry} 
\usepackage[yyyymmdd]{datetime}
\usepackage{algorithm}
\usepackage[noend]{algpseudocode}
\usepackage{lipsum}
\usepackage{setspace}
\usepackage{siunitx}
\usepackage{pdfpages}
\usepackage{tikz}
\usetikzlibrary{positioning}
\usepackage{booktabs,caption}
\usepackage[]{hyperref}
\usepackage{longtable}

\hypersetup{
  colorlinks   = true, 
  urlcolor     = blue, 
  linkcolor    = black, 
  citecolor   = blue 
}
\usepackage[utf8]{inputenc}
 
\usepackage{listings}
\usepackage{color}
\definecolor{codegreen}{rgb}{0,0.6,0}
\definecolor{codegray}{rgb}{0.5,0.5,0.5}
\definecolor{codepurple}{rgb}{0.58,0,0.82}
\definecolor{backcolour}{rgb}{0.95,0.95,0.92}
\lstdefinestyle{mystyle}{
    backgroundcolor=\color{backcolour},   
    commentstyle=\color{codegreen},
    keywordstyle=\color{magenta},
    numberstyle=\tiny\color{codegray},
    stringstyle=\color{codepurple},
    basicstyle=\footnotesize,
    breakatwhitespace=false,         
    breaklines=true,                 
    captionpos=b,                    
    keepspaces=true,                 
    numbers=left,                    
    numbersep=5pt,                  
    showspaces=false,                
    showstringspaces=false,
    showtabs=false,                  
    tabsize=2,
    escapeinside={<@}{@>},
}
 
\usepackage{booktabs} 
\usepackage{siunitx} 
\usepackage{comment}
\usepackage{mathtools} 
\usepackage{gensymb} 
\usepackage{comment}
\usepackage{mhchem} 
\usepackage{fixltx2e} 
\usepackage{bbm}

\usepackage{multirow}

\usepackage{fancyhdr} 
\theoremstyle{definition}

\theoremstyle{definition}

\theoremstyle{remark}

\usepackage{soul} 
\soulregister\cite7
\soulregister\ref7
\soulregister\pageref7
\usepackage{bm}

\usepackage{framed} 

\usepackage{multicol} 

\usepackage{nomencl} 
\usepackage{tikz}
\makenomenclature
\usepackage[resetlabels,labeled]{multibib}

\renewcommand*\nompreamble{\begin{multicols}{2}}

\renewcommand*\nompostamble{\end{multicols}}

\usepackage{amssymb}
\usepackage{pifont}
\definecolor{light-gray}{gray}{0.95}

\journal{a Journal (Under Review)}

\newcolumntype{C}[1]{>{\centering\arraybackslash}m{#1}}

\begin{document}

\begin{sloppypar}

\begin{frontmatter}

\title{\large A Comparative Analysis of Text-to-Image Generative AI Models in Scientific Contexts: A Case Study on Nuclear Power}

\author{Veda Joynt$^{a,\bigstar}$, Jacob Cooper$^{a,\bigstar}$, Naman Bhargava$^{b,\bigstar}$, Katie Vu$^{c,\bigstar}$, O Hwang Kwon$^{a,\bigstar}$, Todd R. Allen$^{a}$, Aditi Verma$^{a}$, Majdi I. Radaideh$^{a,*}$}

\cortext[mycorrespondingauthor]{Corresponding Author: Majdi I. Radaideh (radaideh@umich.edu)}

\address{$^{a}$Department of Nuclear Engineering and Radiological Sciences, University of Michigan, Ann Arbor, MI 48109, United States}
\address{$^{b}$Department of Statistics, University of Michigan, Ann Arbor, MI 48109, United States}
\address{$^{c}$Department of Computer Science and Engineering, University of Michigan, Ann Arbor, MI 48109, United States}

\author{$^{\bigstar}$Authors who have contributed equally to this manuscript}

\begin{abstract}
In this work, we propose and assess the potential of generative artificial intelligence (AI) to generate public engagement around potential clean energy sources. Such an application could increase energy literacy  -- an awareness of low-carbon energy sources among the public therefore leading to increased participation in decision-making about the future of energy systems. We explore the use of generative AI to communicate technical information about low-carbon energy sources to the general public, specifically in the realm of nuclear energy. We explored 20 AI-powered text-to-image generators and compared their individual performances on general and scientific nuclear-related prompts. Of these models, DALL-E, DreamStudio, and Craiyon demonstrated promising performance in generating relevant images from general-level text related to nuclear topics. However, these models fall short in three crucial ways: (1) they fail to accurately represent technical details of energy systems; (2) they reproduce existing biases surrounding gender and work in the energy sector; and (3) they fail to accurately represent indigenous landscapes -- which have historically been sites of resource extraction and waste deposition for energy industries. This work is performed to motivate the development of specialized generative tools and their captions to improve energy literacy and effectively engage the public with low-carbon energy sources. 
\end{abstract}


\end{frontmatter}

\setstretch{1.3}

\section{Introduction}
\label{sec:intro}
There exists a growing global consensus on the need to orchestrate energy transitions to avert the worst effects of climate change. As a result, significant efforts are being made around the world to transform our energy systems, and, as part of this process, engage with communities to increase public awareness about various clean energy options \cite{faria2015new}. These efforts have resulted in increased understanding, and acceptance of solar and wind energy technologies.  [deleted what was previously here. Instead, insert a few sentences on energy literacy]

The emergence of state-of-the-art Text-to-Image Generative Artificial Intelligence (AI) Models, such as DALL-E, could potentially be used spread awareness of these systems by generating technical diagrams. This project explores the performance of generative AI models in creating realistic scientific images to motivate public engagement and increase public awareness of these unknown clean energy solutions. In this paper, Generative AI Models are defined to be "models that create images from another and from different types of data including but not limited to text, scene, graph and object layout" \cite{elasri2022image}.

In 2021, OpenAI's Text-to-Image Generative AI Model DALL-E was made public. Since then, there has been a growing interest in exploring the potential of these models for creative and technical applications. However, the demand for generative AI models emerged before 2021 \cite{sapkota2023harnessing, vartiainen2023using, mccrum2008realistic_PANGU}.  Various motivations, like data augmentation, drove generative AI model development.  Previous studies in this field have utilized generative AI models for educational purposes, such as clothing and accessory image generation for craft education, crop image generation in agriculture settings, in architecture and urban design, as well as novel art generation \cite{sapkota2023harnessing, vartiainen2023using}. In the fashion industry, generative AI models are used to improve designer efficiency; for example, Yan et al. \cite{9693190} created a training data set of 115,584 pairs of fashionable items, which was used to test generative text-to-image AI performance.

\begin{figure}[!h]
    \centering
    \includegraphics[width=\textwidth]{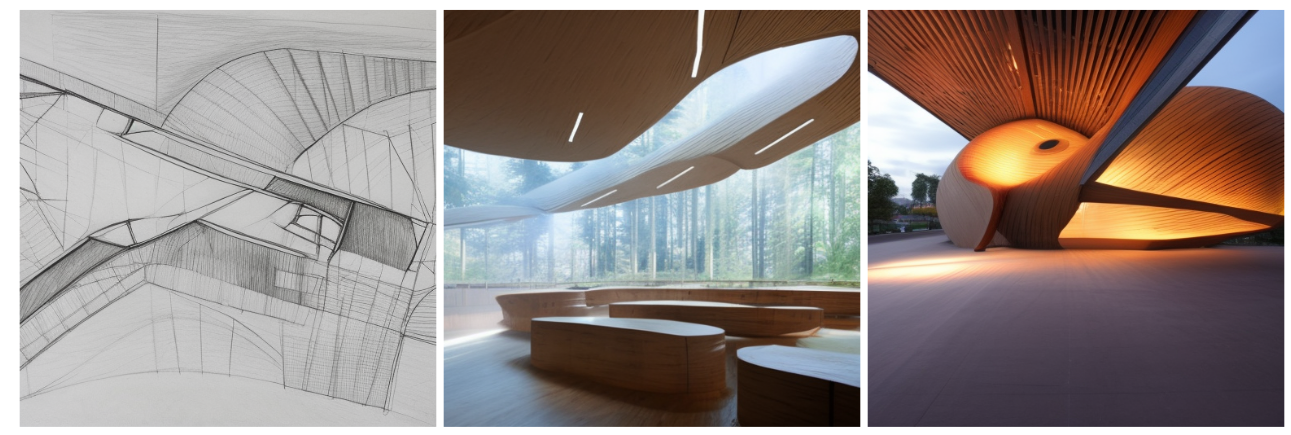}
    \caption{The participants' floorplans, interior views, and facade materials which are voted as the best designs in study \cite{paananen2023using_architecture}}
    \label{fig:architecture}
\end{figure}

In 2008, McCrum et al. \cite{mccrum2008realistic_PANGU} used an generative AI models to create realistic images to simulate Martian exploration robots in the software Planet and Asteroid Natural Scene Generation Utility (PANGU). More recently, there has been a dynamic movement towards utilizing generative AI models to create images to improve the quantity and diversity of training data in predictive medical diagnostic programs \cite{akrout2023diffusion_skin}. To tackle the challenges arising in acquiring massive data due to patient privacy concerns, Akrout et al \cite{akrout2023diffusion_skin} employed such generative AI models for data augmentation. The commonality among the presented studies \cite{mccrum2008realistic_PANGU,akrout2023diffusion_skin} above is that, when data is relatively scarce, generative AI models can augment data to improve the performance of machine learning classification algorithms by reprocessing existing data. Generative AI models have been applied to areas beyond data augmentation. These areas include generation of novel artworks and the production of visual materials for communication \cite{paananen2023using_architecture,seneviratne2022dalle}. Recently, highly sophisticated and imaginative generative AI models have even been used in the field of architectural design. In Paananen et al. \cite{paananen2023using_architecture}, students were tasked to design a culture center in a small island using generative AI models, namely DALL-E, StableDiffusion, and Midjourney. Figure \ref{fig:architecture} illustrates one of the standout works identified as the best in \cite{paananen2023using_architecture}.

While the previously mentioned applications have employed generative AI models in a positive context, it is important to recognize that there are also negative implications and ethical concerns of AI image generators. A concern with generative AI images is copyright violations. Images are extracted from search engines, such as Google, to train a generative AI model. Since many of these images are protected by copyright, the resulting image produced by the generative AI models may breach copyright law as these images are trained without direct consent of the creators \cite{10139768}. Generative AI models could also intentionally be used to generate images that portray a false representation of reality or contain disinformation. Such works like deepfakes could be used to damage reputations, blackmailing individuals' for monetary benefits, inciting political or religious unrest by targeting politicians or religious scholars with fake videos/speeches dissemination, as well as spread disinformation about current events \cite{masood2023deepfakes}. Additionally, images produced by generative AI could additionally reflect and perpetuate stereotypical, racist, discriminatory, and sexist ideologies. For example, Buolamwini and Gebru \cite{buolamwini2018gender} reported that two facial generative AI training data sets, IJB-A and Adience, are composed of 79.6\% and 86.2\% lighter-skinned subjects, respectively, rather than darker-skinned subjects. It was also found that darker-skinned females are the most likely to be incorrectly classified, with a classification error rate of 34.7\% \cite{buolamwini2018gender}.

While there have been many studies highlighting the use of generative AI models in areas such as medical diagnostics, robotic motion planning, fashion, art, architecture and urban design, we discovered there are a lack of studies that utilize generative AI models to address climate change-related problems for most clean energy sources from a technical and policy perspective. Of the sources evaluated in our literature review, only one study \cite{qadri2023ai} performed a community-centered study of cultural limitations of generative AI models. However, this study was specifically performed in the regime of the South Asian context to study the impact of global and regional power inequities. Another study \cite{wang2018public} emphasized that there exists a need to incorporate visual images alongside written language to impact public perceptions of climate change. However, generative AI was not explicitly employed in the study. Due to this lack of prior investigation, our research team was interested in determining whether generative AI models can produce technically accurate image that reflects the given prompt even in specialized and technically sophisticated engineering-oriented image scenarios. Furthermore, previous studies primarily relied upon widely recognized tools, such as DALL-E, Stable Diffusion, and Midjourney. For example, Sapkota et al. \cite{sapkota2023harnessing} used MidJourney and Vartiainen et al. \cite{vartiainen2023using} used Dall-E 2. Recently, a plethora of models have been introduced beyond the three aforementioned models. Our research team has taken the initiative to directly engage with these models, evaluating their pros and cons in the process. In this paper, our team conducted a case study on generative AI models to test performance and accuracy related to nuclear energy prompts. We analyzed 20 different generative AI models, with an emphasis on the tools with an accessible Python API. We then selected the top 3 performing models among 20 models based on accessibility, image quality, accurate portrayal of prompts, process time, and cost. For this process, we selected prompts related to different nuclear power plant components and processes occurring within a power plant, ran these prompts through the top 3 generative AI tools, applied prompt engineering to enhance the generators ability to create images that reflect the given prompt as well as increase the technical accuracy of the images. Finally, we analyzed the performance of these tools in regard to their technical accuracy in depicting different nuclear engineering components such as radiation shielding of a nuclear reactor, the primary side of a pressurized water reactor, etc.

\textit{The work presented in this paper is novel for several reasons}. First, the majority of these generative AI tools began their maturation process a few years ago, with a restricted amount of literature and analysis available regarding their technical accuracy in a scientific context. Second, our literature survey indicates a minimal application of generative AI for generating images to foster community engagement and enhance public perspectives on climate change. Third, this study assesses the robustness of current state-of-the-art generative AI models and assess the necessity of specialized generative AI tools within specific disciplines where models are trained on discipline-focused images and text captions. Such applications to nuclear energy include nuclear fuel rod fabrication, proper waste management images, and nuclear reactor designs.

This rest of this paper is organized as follows: Section \ref{sec:background} presents the concepts behind how generative AI models work and their primary features. In section \ref{sec:method}, we compare all testing generative AI models according to several factors by discussing their advantages and disadvantages. Section \ref{sec:results} presents the generative AI results for the top 3 generative AI models based on similar prompts. The conclusions of this work and the potential opportunities for future work are highlighted in section \ref{sec:conc}. 

\section{Generative AI}
\label{sec:background}

\subsection{Generative AI Concepts}

\begin{figure}[!h]
    \centering
    \includegraphics[width=\textwidth]{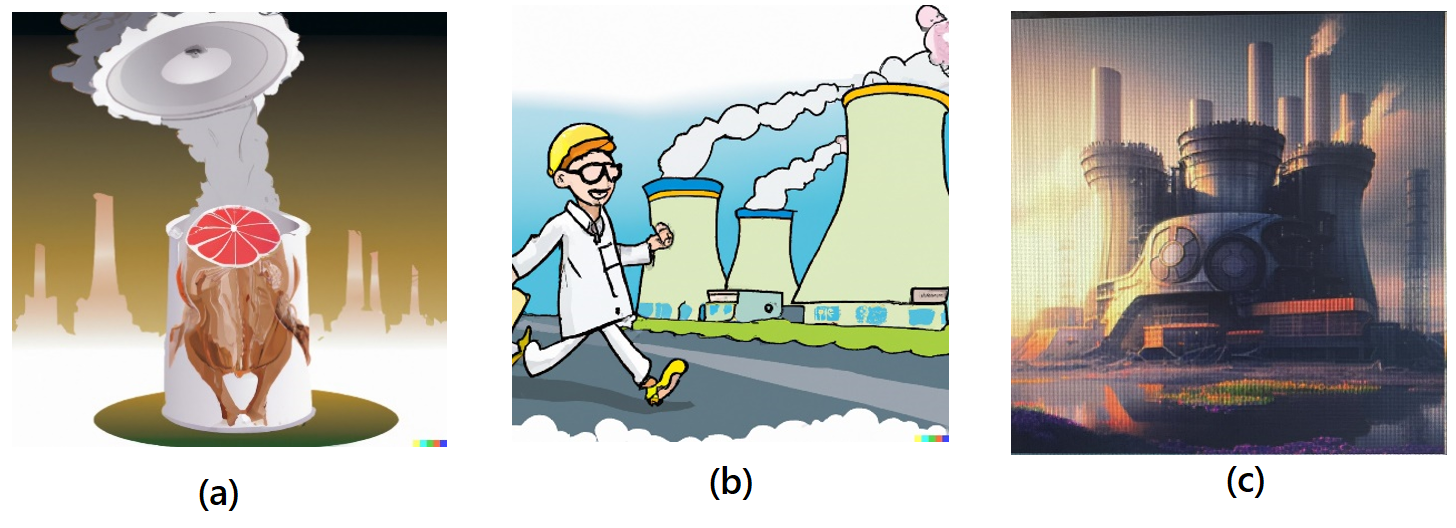}
    \caption{a. A turkey cooked on a nuclear power steam (DALL-E 2), b. A happy man working at a nuclear power plant without any risks (DALL-E 2), and c. nuclear power plant (Midjourney)} 
    \label{fig:example of text to image}
\end{figure}

Generative text-to-image AI models are a subset of generative AI models that take text input and create an image based on the input description. Figure \ref{fig:example of text to image} illustrates images generated by various models using text as a basis. Figures \ref{fig:example of text to image}(a) and \ref{fig:example of text to image}(b) were generated using DALL-E, while Figure \ref{fig:example of text to image}(c) was produced through Midjourney. Generative AI models can create logical as well unusual images that would be difficult to find elsewhere, such as a turkey inside a nuclear cooling tower in Figure \ref{fig:example of text to image}. 

\begin{figure}[!h]
    \centering
    \includegraphics[width=1.0\textwidth]{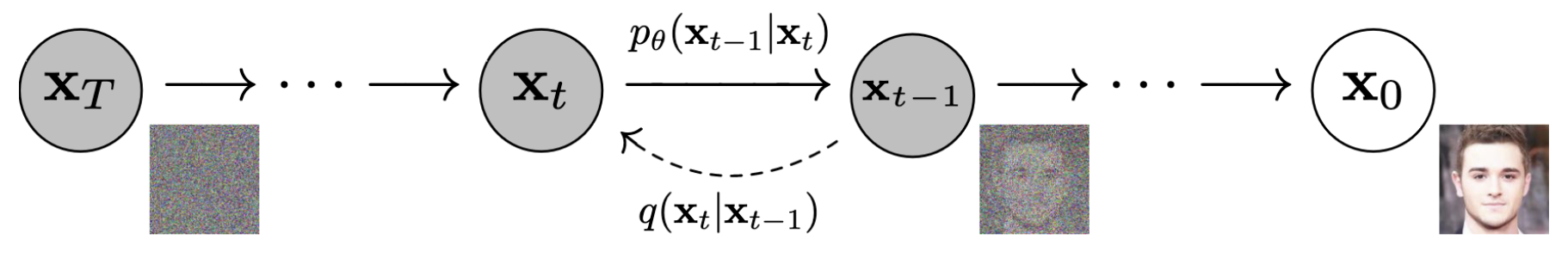}
    \caption{Diffusion Model Flow \cite{ho2020denoising}} 
    \label{fig:Transformer}
\end{figure}

As evident from Figure \ref{fig:example of text to image},  generative AI models process the captions provided by the user and reproduce corresponding images. Interestingly, both DALL-E and Midjourney generated images of cooling towers in response to the text "nuclear power plant." This suggests that these models have been pre-trained to associate the text "nuclear power plant" with the concept of a cooling tower, likely because cooling towers are often the most visually prominent aspect of images of nuclear reactors. When generative AI models produce images of cooling towers, they are capturing an important feature of a nuclear plant but failing to depict other important features such as the reactor system itself. This is one of the gaps that we found in this work. The training process of text-to-image generative AI models is briefly described next:

\begin{enumerate}
    \item \textit{Training Concept}: Generative AI models use a pre-trained data set of images that link natural language to an image. A popular pre-trained deep learning model is Contrastive Language Image Pretraining (CLIP), developed by OpenAI; CLIP was trained on 400 million images with text \cite{schuhmann2021laion}. CLIP learns the weight of how much a caption relates to a given image. CLIP follows this workflow: Images and text captions are passed through encoders, which map all objects to a $m$-dimensional space \cite{radford2021learning}. The cosine similarity is taken from the text caption and image. Ideally, we should maximize the cosine similarity between $N$ correct encoded image and text caption pairs, while minimizing the cosine similarity between $N^2 - N$ incorrect encoded image and text caption pairs \cite{radford2021learning}. By comparison, other examples of models that use contrastive learning are ALIGN and CLOOB. 

    \item \textit{Decoding and Transformer Models}: CLIP learns both an image and text encoding; Radford et al. \cite{radford2021learning} used autoregressive models and diffusion models to perform the mapping of text caption encoding to image encoding for each image. The researchers found that both models produced similar results and diffusion models were computationally less intensive \cite{radford2021learning}. Next, the diffusion model (also known as diffusion prior) is utilized to map the text caption encoding to image encoding, displayed in Figure \ref{fig:Transformer}. After CLIP receives the encoded image and caption data, the generative AI tool (e.g., DALL-E) needs to reverse this to generate an image; it uses a diffusion model (also known as diffusion posterior) to decode the CLIP encoded data. Inspired by the principles of thermodynamics, diffusion models are text-to-image models that add Gaussian noise to data and then reverse the diffusion process to restore clarity to the Gaussian blurred images \cite{ho2020denoising}. DALL-E uses GLIDE, a transformer model by Open-AI, to decode image and caption data from CLIP. For CLIP, it encodes the caption string into tokens, takes tokens and inputs that to transformer, generates output tokens, conditions the final token for embedding, and then combines projection of final token embedding to additional context \cite{nichol2021glide}.  
\end{enumerate}

\begin{figure}[!h]
    \centering
    \includegraphics[width=\textwidth]{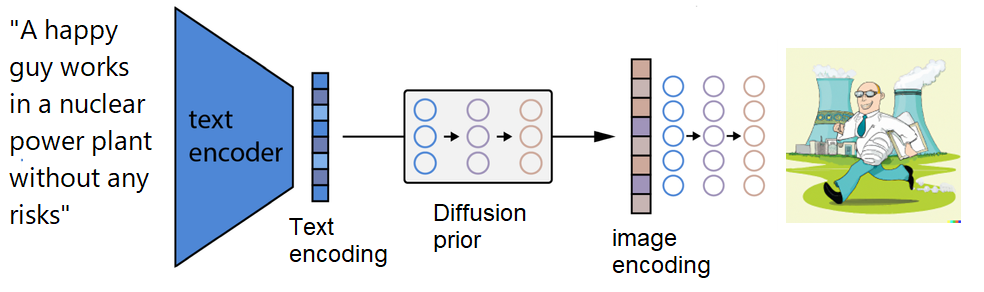}
    \caption{A schematic of how text-to-image generative AI tools work (modified from \cite{ramesh2022hierarchical})} 
    \label{fig:generative_graphical_models}
\end{figure}

Figure \ref{fig:generative_graphical_models} shows the format of how text-to-image generative AI works. A user inputs a text prompt, CLIP's encoder maps this into $m$-dimensional space, the diffusion prior maps the CLIP text encoding to the corresponding CLIP image encoding, then the GLIDE model will use reverse-Diffusion to map the CLIP text and image encoding to generate images based on the inputted text description \cite{nichol2021glide}.

\subsection{Generative AI Features}

The prompt is the caption that creates an image; generative AI tools rely on a prompt to take the user's intention to generate an image. Some generative AI tools such as Canva's text-to-image generation service has a graphical user interface (GUI) that allows a user to input prompts, and generate an image based on that prompt. Other generative AI tools have either free or paid API access, where a user can input a prompt into a Python script. When employing generative AI through a GUI, while the usage is intuitive, it may not be optimal for generating a large volume of images using extensive prompts. Consequently, the most ideal scenario arises when both GUI and API are concurrently available.

In order to fully take advantage of text-to-image generative AI models, we looked for models that supported a text input prompt, inpainting, outpainting, model training, and image-to-image editing. Each of these terms are described below. 

\begin{figure}[!h]
    \centering
    \includegraphics[width=0.7\textwidth]{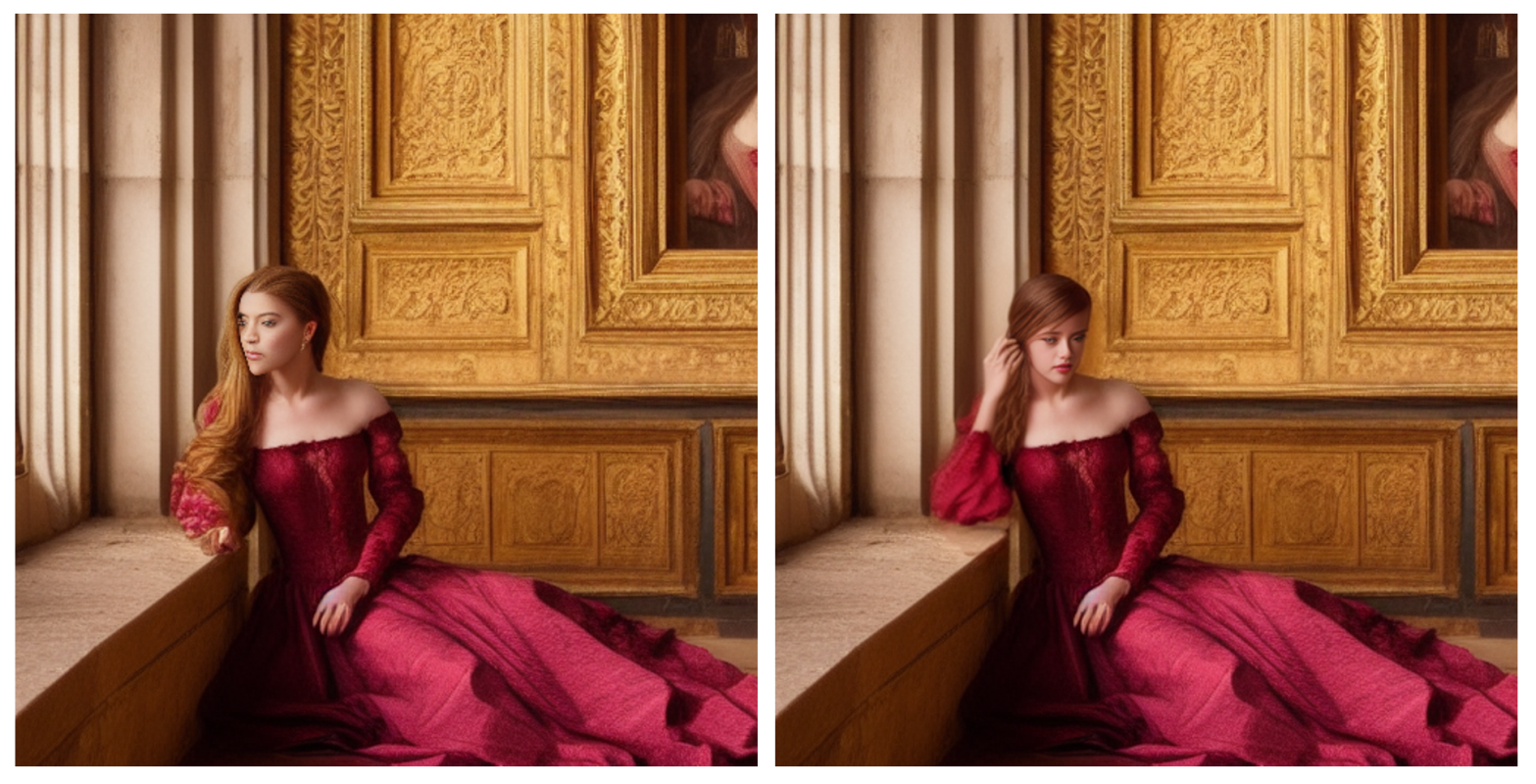}
    \caption{Inpainting from Stable Diffusion: Rotating Head and Hand \cite{inpainting}} 
    \label{fig:inpaintingex} 
\end{figure}

Inpainting is a tool used to take missing or unknown parts of an image and use AI to generate this unknown region \cite{zhang2022gan}. Generative AI models are trained on an extensive set of images; inpainting takes its trained data set to replace specific parts of an image. Inpainting is most commonly used for the removal of unwanted objects, image restoration, and image editing \cite{doria2012filling}. Figure \ref{fig:inpaintingex} shows an example of inpainting from StableDiffusion. 

\begin{figure}[!h]
    \centering
    \includegraphics[width=\textwidth]{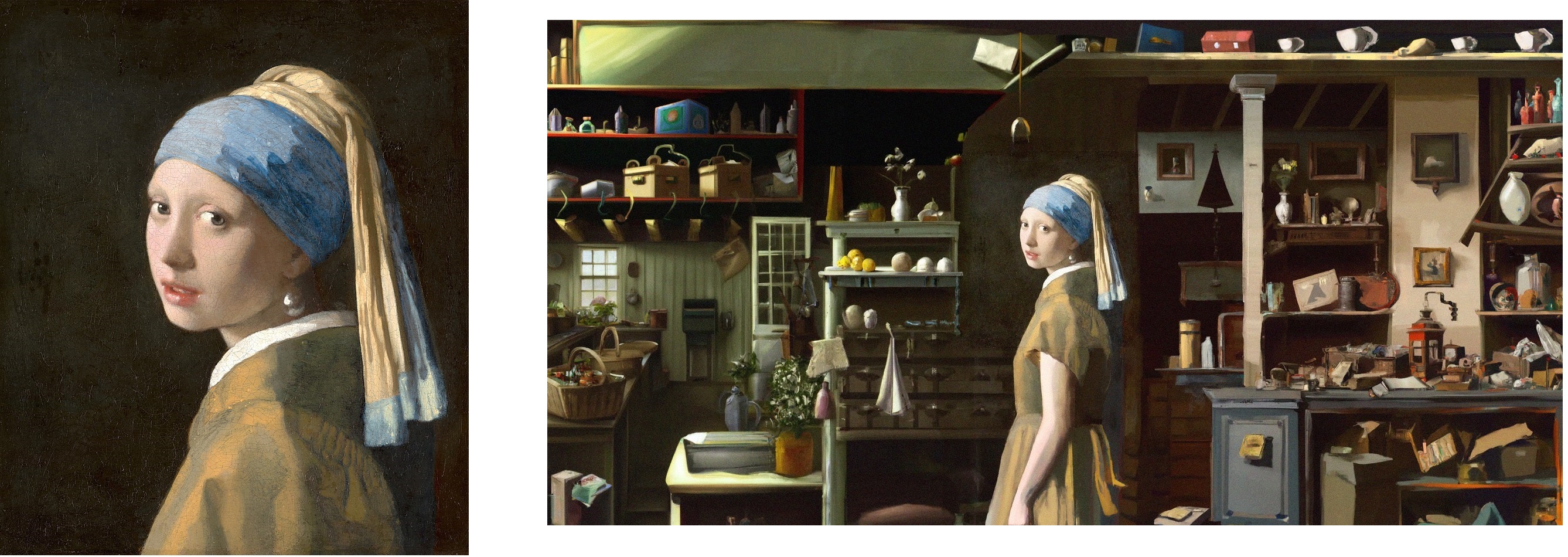}
    \caption{Girl with a Pearl Earring (left) and DALL-E 2 outpainting of \textit{``Girl with a Pearl Earring''} (right) \cite{outpainting}} 
    \label{fig:outpainting}    
\end{figure}

Outpainting is the opposite of inpainting; outpainting is a tool used to extend the borders to add additional parts to the image using AI \cite{xiao2020image}. Outpainting can be used to change the aspect ratio of an image and extend borders to an image. Figure \ref{fig:outpainting} shows an example of outpainting using DALL-E model. 

\begin{figure}[!h]
    \centering
    \includegraphics[width=0.7\textwidth]{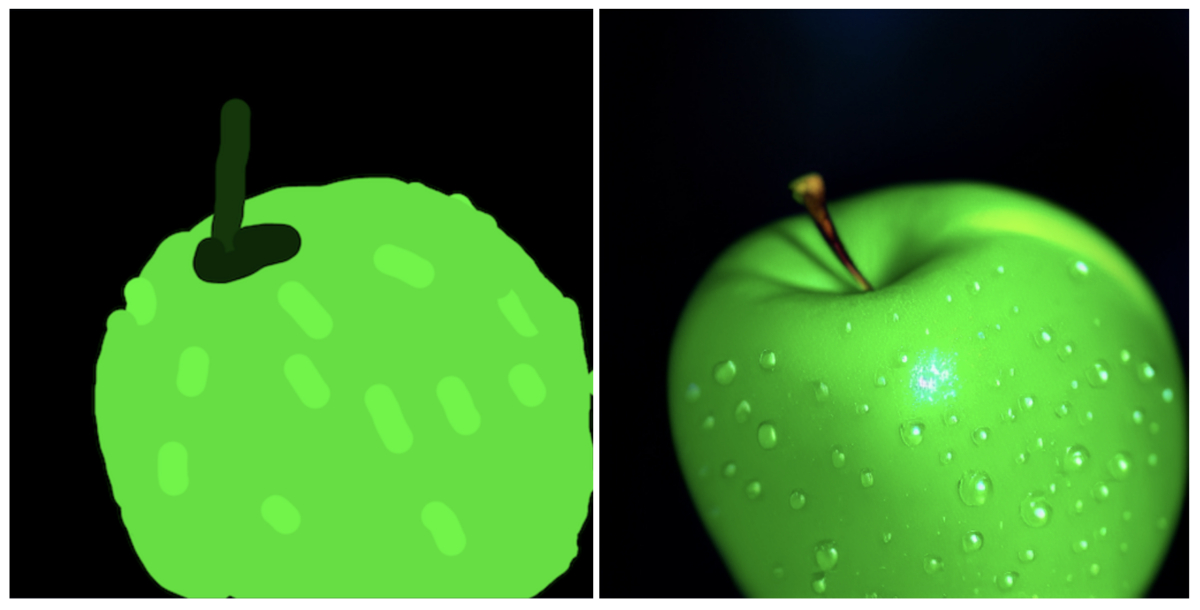}
    \caption{Image to Image of Apple Sketch from Stable Diffusion \cite{i2i}} 
    \label{fig:i2i}
\end{figure}

Image-to-image models take an image for an input, and allow specific edits to be made to yield a fine-tuned output \cite{brooks2023instructpix2pix}. Commonly, image-to-image models will allow for style changes, altering resolution, and generation of high-quality images from low-quality images. Figure \ref{fig:i2i} shows a high-quality creation of an apple from a basic sketch performed using image-to-image technology in StableDiffusion. 

"Model training feature" refers to the training of a machine learning model, usually a neural network, with a sample of image-caption pair. The generative AI will then use an input sample of the images for training data and output several images following a prompt \cite{kumari2022ensembling}. For instance, to develop a specialized generative AI tailored to nuclear power, one can train the model using captions such as "nuclear power plant" accompanied by a variety of images from different nuclear power plants. Ideally speaking, through this training, the AI becomes capable of producing more realistic and accurate images when prompted with content related to "nuclear".

\section{Methodology}
\label{sec:method}

\subsection{Comparison of Generative AI Models}
We tested 20 total text-to-image generative AI tools, each with varying results shown in Table \ref{tab:successful} for the tools with promising performance and in Table \ref{tab:unsuccessful} for the tools with poor performance. In our initial evaluation, we first identified which tools had API access. Tools that did not have API access were then removed such as Nightafe, Fotor AI and Artbreeder.  Additionally, tools such as DreamStudio that used the same API as another model (Stable Diffusion) were also removed. Then we narrowed down our tools based on the ability to generate images. Parti and Google Brain Images were eliminated because they are not available to the public. DeepAI was similarly eliminated due to the availability of only paid subscription services. Other text-to-image generative AI tools such as StackGan++ and CLIP required training data in order to generate images, and thus were also eliminated. Midjourney is a popular text-to-image generative AI model, but we decided against using Midjourney since a Discord account is required to access the API, and due to increased usage, the servers were generally unavailable. We then reviewed the commercial rights of these programs and found that Leonardo.Ai did not support commercial use and was thus removed. Starryai, Picsart, Kapwing, Writesonic additionally had poor technical image quality when tested on basic prompts including "Display radioactive nuclear waste" and "China and Nuclear." These prompts were selected due to their simplicity; models were removed if provided poor technical details following a basic nuclear prompt (i.e. faces getting morphed in a nuclear setting, not displaying cooling towers). A summary of the basic prompts, along with prompt results are provided in Table \ref{tab:promisingnucresults}. Of the remaining systems, DALL-E 2 and Stable Diffusion both had paid subscriptions; however, they were chosen due to their capabilities of inpainting/outpainting, image-to-image editing, and good performance in image generation. In contrast, Canva and Craiyon both have free subscriptions but no inpainting/outpainting and image-to-image editing.  However, Canva had a very long generation time for images, when compared to all of the other models and was thus removed.

\begin{table}[!h]
\caption{Successful Generative AI Models}
\centering
\label{tab:successful}
\begin{tabular}{l|l|l|l}
\hline
Generative AI Model         & Developer        & Pros                                                                                                                           & Cons                                                                                                                           \\ \hline
DALL-E 2                    & OpenAI           & \begin{tabular}[c]{@{}l@{}}Inpainting and outpainting\\ Commercial rights ownership\\ API access\end{tabular}                  & \begin{tabular}[c]{@{}l@{}}Costly credit system\\ Some low quality images\\ Paid API\end{tabular}                              \\ \hline
Stable Diffusion            & CompVis Group    & \begin{tabular}[c]{@{}l@{}}Inpainting and outpainting\\ Commercial rights ownership\\ Negative prompt box\end{tabular}         & \begin{tabular}[c]{@{}l@{}}Costly credit system\\ No style and variation\end{tabular}                               \\ \hline
Leonardo.Ai                 & Leonardo.Ai      & \begin{tabular}[c]{@{}l@{}}Inpainting and outpainting\\ Several editing options\end{tabular}                                   & \begin{tabular}[c]{@{}l@{}}Leonardo.Ai owns image rights\\Poor technical results\end{tabular}                     \\ \hline
Craiyon & Boris Dayma      & \begin{tabular}[c]{@{}l@{}}Free text to image works well\\ Has free API\\ High quality images\end{tabular} & \begin{tabular}[c]{@{}l@{}}Watermark on free images\\ Advertisement supported\\ Commercial use is unclear\end{tabular} \\ \hline
DreamStudio                 & Cristofer Adrian & \begin{tabular}[c]{@{}l@{}}Stable Diffussion's API\\ Cleaner interface and open-source\end{tabular}                            & Costly credit system                                                                                                           \\ \hline
Photosonic                  & Writesonic       & \begin{tabular}[c]{@{}l@{}}Comercial rights ownership\\ Generates 2D and 3D images\end{tabular}                                & \begin{tabular}[c]{@{}l@{}}API is paywalled\\Poor quality images\\ No inpainting and outpainting\end{tabular}   \\ \hline
Jasper Art                  & Jasper.ai        & \begin{tabular}[c]{@{}l@{}}Easy to use interface\\ No credit limit/free to use\end{tabular}                                    & Poor technical image results                                                                                                   \\ \hline
Canva AI                    & Canva            & \begin{tabular}[c]{@{}l@{}}Easy to use interface\\ No credit limit/free to use\end{tabular}                                    & \begin{tabular}[c]{@{}l@{}}Long image generation time\\ Poor technical results\end{tabular}                          \\ \hline
\end{tabular}
\end{table}

\begin{table}[!h]
\caption{Unsuccessful Generative AI Models}
\centering
\label{tab:unsuccessful}
\begin{tabular}{l|l|l|l|l}
\hline
Generative AI Model & Developer     & Pros                                                                                                                      & Cons & Success                                                                                          \\ \hline
Midjourney          & Midjourney    & \begin{tabular}[c]{@{}l@{}}Inpainting and outpainting\\ Commercial rights ownership\\ High resolution images\end{tabular} & \begin{tabular}[c]{@{}l@{}}Issues with free trial\\ Requires Discord account\end{tabular} & yes         \\ \hline
starryai            & Mo Kahn       & Easy to use interface                                                                                                     & Poor image quality

\\ \hline
NightCafe           & Angus Russell & Free basic plan                                                                                                           & \begin{tabular}[c]{@{}l@{}}No inpainting and outpainting\\ No API\\ No editing images\end{tabular} & yes \\ \hline
Artbreeder          & Joel Simon    & Allows image editing                                                                                                      & \begin{tabular}[c]{@{}l@{}}Costly credit system\\ Poor image quality\end{tabular}  & yes                  \\ \hline
Fotor AI            & Fotor         & \begin{tabular}[c]{@{}l@{}}Free \\ High quality images\end{tabular}                                                       & \begin{tabular}[c]{@{}l@{}}10 image limit per day\\ No API\end{tabular}     & yes                         \\ \hline
Google Brain Imagen & Google        & High quality images                                                                                                       & Not available to public                                                                           & yes  \\ \hline
CLIP                & OpenAI        & \begin{tabular}[c]{@{}l@{}}Free \\ Open source\\ Inpainting and outpainting\end{tabular}                                 & \begin{tabular}[c]{@{}l@{}}Requires training data\\ Replaced by DALL-E\end{tabular}              & no   \\ \hline
StackGan++          & Han Zhang     & \begin{tabular}[c]{@{}l@{}}Free \\ Open source\\ Inpainting and outpainting\end{tabular}                                 & \begin{tabular}[c]{@{}l@{}}Requires training data\\ Poor image quality\end{tabular}              & yes   \\ \hline
Parti               & Google        & Text to image                                                                                                             & Not available to public                                                                           & no \\ \hline
Picsart             & Picsart       & \begin{tabular}[c]{@{}l@{}}Free \\ Text to image\end{tabular}                                                             & Poor image quality                                                                                 & yes\\ \hline
Kapwing             & Kapwing       & Commercial rights ownership                                                                                               & \begin{tabular}[c]{@{}l@{}}Free for several images\\ Poor image quality\end{tabular}              & yes  \\ \hline
DeepAI              & DeepAI        & High quality images                                                                                                       & \begin{tabular}[c]{@{}l@{}}Paid subscription\\ Poor technical results\end{tabular}              & yes    \\ \hline
\end{tabular}
\end{table}

\subsection{Prompt Engineering}

Prompt Engineering refers to optimizing the prompt (text input to models) for generating desired images from text-to-image generative AI models. Prompt Engineering can help in achieving the desired result from a pre-trained model, reducing the need of computational resources and knowledge to fine-tune these models for different tasks \cite{gu2023systematic}. Apart from text-to-image models, this method has been applied to other generative models as well, like GPT-3 and ChatGPT, which are text-to-text generative AI models. 

Prompt Engineering is an iterative process and helps in efficient interaction with the latent space of generative models. Researchers have identified and classified different type of keywords to produce images closer to desired results   \cite{oppenlaender2022taxonomy}. Certain types of keywords, such as \textit{'hyperrealistic'}, \textit{'oil on canvas'}, \textit{'abstract painting'},\textit{'in the style of a cartoon'},  are especially useful in directing the style of the image, as displayed in Table \ref{tab:imgstyle}. Therefore, such keywords have been used in this study as well to generate images closer to real-life.

\begin{figure}[!h]
    \centering
    \includegraphics[width=7cm, height=12cm]{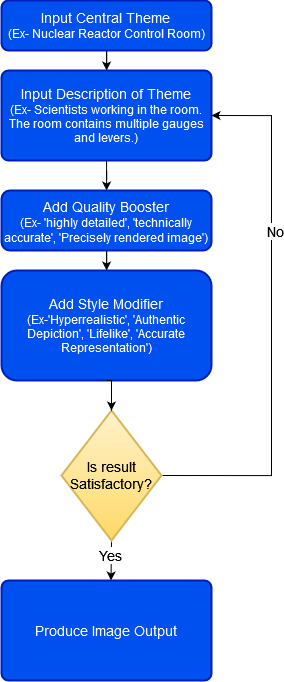}
    \caption{Prompt Engineering Algorithm} 
    \label{fig:pealgo}
\end{figure}

One of the main characteristics identified by the authors for images generated related to nuclear energy was that images should look realistic in order to avoid exaggeration. This exaggeration can occur in the images due to the artistic nature of these models (see for example sample results in Table \ref{tab:craiyonpe}). Further, the images should be detailed to capture the intricacies of different components, especially in case of technical designs. To ensure that these characteristics are reproduced in the images generated by the generative models, we decided to include style modifiers keywords and quality booster keywords in the prompts \cite{oppenlaender2022taxonomy}, ensuring realistic flair and high detailing of the images. Further, to improve the image quality, an additional description of the theme regarding the visual appearance of the subject was appended to the prompt. Figure \ref{fig:pealgo} displays the flowchart for prompt engineering we adopted in this study. The method is implemented in an iterative manner, changing keywords and descriptions associated with the prompt to get as realistic results as possible.  

\begin{table}[!h]
\caption{Dictating image style of Nuclear Cooling Tower with prompt engineering}
\label{tab:imgstyle}%
\centering
    \begin{tabular}{m{2cm}|m{3.5cm}|m{3.5cm}|m{3.5cm}|m{3.5cm}}
    \toprule
    Style modifier & Oil on canvas  & In the style of a cartoon  & Hyperrealistic  & Abstract painting \\
    \midrule
     Image& \includegraphics[width=3cm, height=3cm]{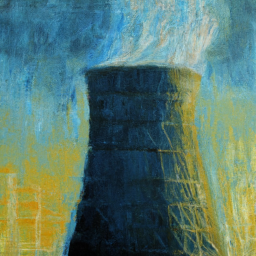} & \includegraphics[width=3cm, height=3cm]{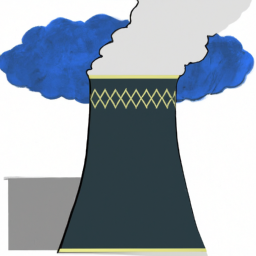} & \includegraphics[width=3cm, height=3cm]{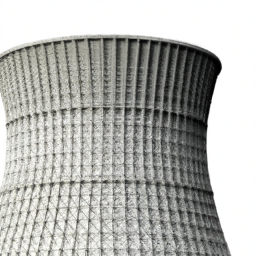}
     & \includegraphics[width=3cm, height=3cm]{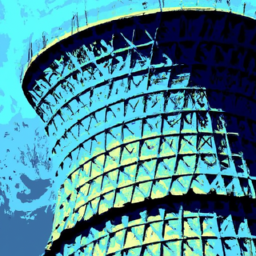}\\
     
    \bottomrule
    \end{tabular}%
\end{table}

\section{Results and Discussions}
\label{sec:results}

Of the 20 AI models explored, we narrowed our focus to three models based on access to API, cost, successful generation of images, and the accurate portrayal of prompts. As our focus in this study is generating high-quality images that accurately illustrate the prompts, we focused our attention on DALL-E, Craiyon, and DreamStudio. Despite the costly credit system of DALL-E and DreamStudio, the tool produces high-quality images in addition to inpainting, outpainting, and image-to-image editing. We also chose Craiyon for optional cost expenses but high-quality image generation.

\subsection{Results for General Prompts} 
\label{sec:Results for General Prompts}

We tested the narrowed pool of 3 generative AI models with 10 prompts selected from initially 36 prompts to evaluate the quality of images, however, for brevity, we have demonstrated two samples in Table \ref{tab:generalresults}. All the AI-powered generator models gave multiple image outputs for a single prompt, out of which the image which portrayed the prompt with highest technical accuracy was chosen. 

In our first prompt, we asked DALL-E 2, DreamStudio, and Craiyon to produce a "High quality image of bunnies in a field." The prompt was selected to demonstrate a commonly viewed nature scene, where surroundings of grass and trees are similar to the surroundings of a nuclear reactor. This produced similar results among the three, each with grass and varying bunny colors. Each bunny appears to be accurate, correctly depicting the ears, head, and body shape. DALL-E 2 produced the most realistic image, and this appears to be a cottontail bunny. This generative AI tool produced extremely realistic grass; however, DALL-E 2 only produced one bunny, when asked to produce "bunnies." DreamStudio produced two bunnies that look realistic. The grass appears to be over-saturated and the bunnies' coloring appears slightly off and looks a little ``cartoonish'' (as interpreted by members of the research team); however, still produced a technically accurate result of bunnies. Craiyon produced two bunnies that appear physically accurate. The grass is out-of-focus and does not look as realistic compared to DALL-E 2 and DreamStudio.

In our second prompt as shown on the Table \ref{tab:generalresults}, we compared the three AI tools with the prompt ``An oil painting of Michigan sand dunes.'' When testing these models, we generated four image outputs for a single prompt, out of which the image which portrayed the prompt with highest technical accuracy was chosen. From these tests we observed that, DALL-E 2 created an image that most resembles an oil painting. It has an accurate depiction of sand, sky, and sea grass. This generative AI tool also does an excellent job at shadows. In comparison, DreamStudio did not necessarily create an oil painting, but did create an image resembling qualities of a painting, such as the appearance of brush strokes and watercolor themes. It correctly depicted sand dunes and sea grass. Craiyon produced a realistic image that we would not consider as an oil painting. The shadows appear to be consistent from a light source relative to the left side of the image. Craiyon accurately generated a large body of water in front of the sand dunes, presumably the great lakes. It accurately depicts sand dunes with a lot of sea grass by the water. 
Overall, it appears that these generative AI models produce accurate details for prompts describing the natural environment.

\begin{table}[!h]
\caption{Image generated from ``general life'' topics}
\label{tab:generalresults}%
\centering
    \begin{tabular}{m{2.7cm}|m{4.2cm}|m{4.2cm}|m{4.2cm}}
    \toprule
    Prompts & DALL-E 2 & DreamStudio & Craiyon \\
    \midrule
     "High quality image of bunnies in a field"& \includegraphics[width=\linewidth]{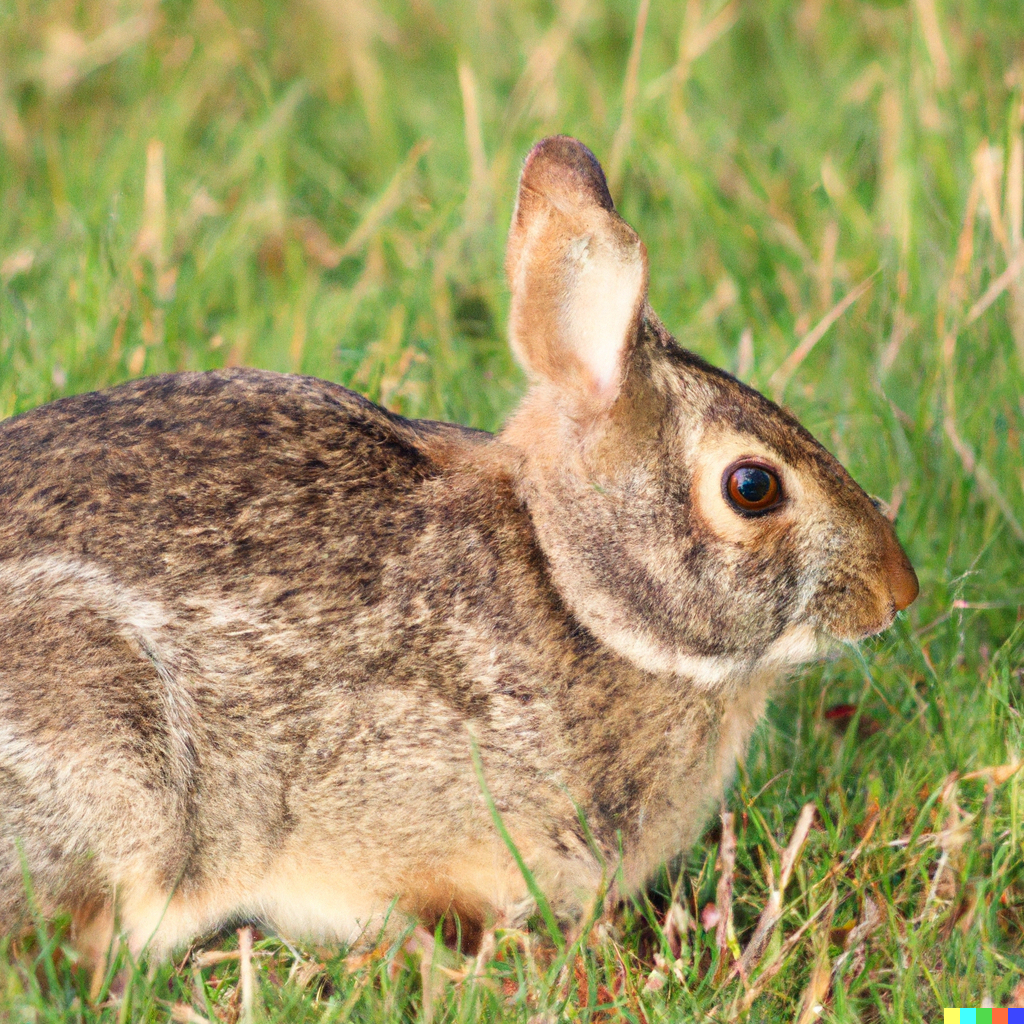} & \includegraphics[width=\linewidth]{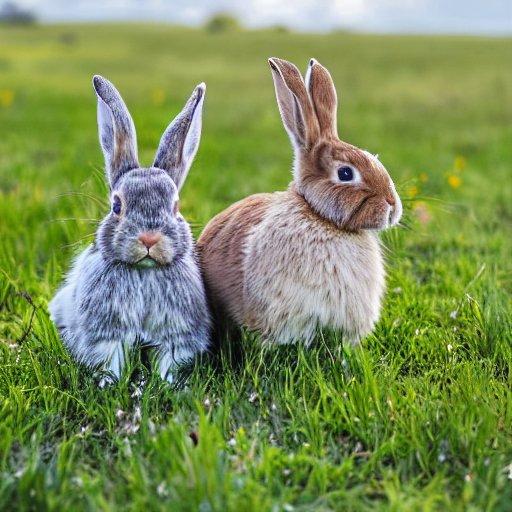} & \includegraphics[width=\linewidth]{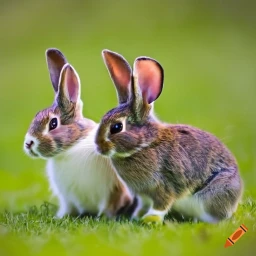} \\
    \midrule
     "An oil painting of Michigan sand dunes"& \includegraphics[width=\linewidth]{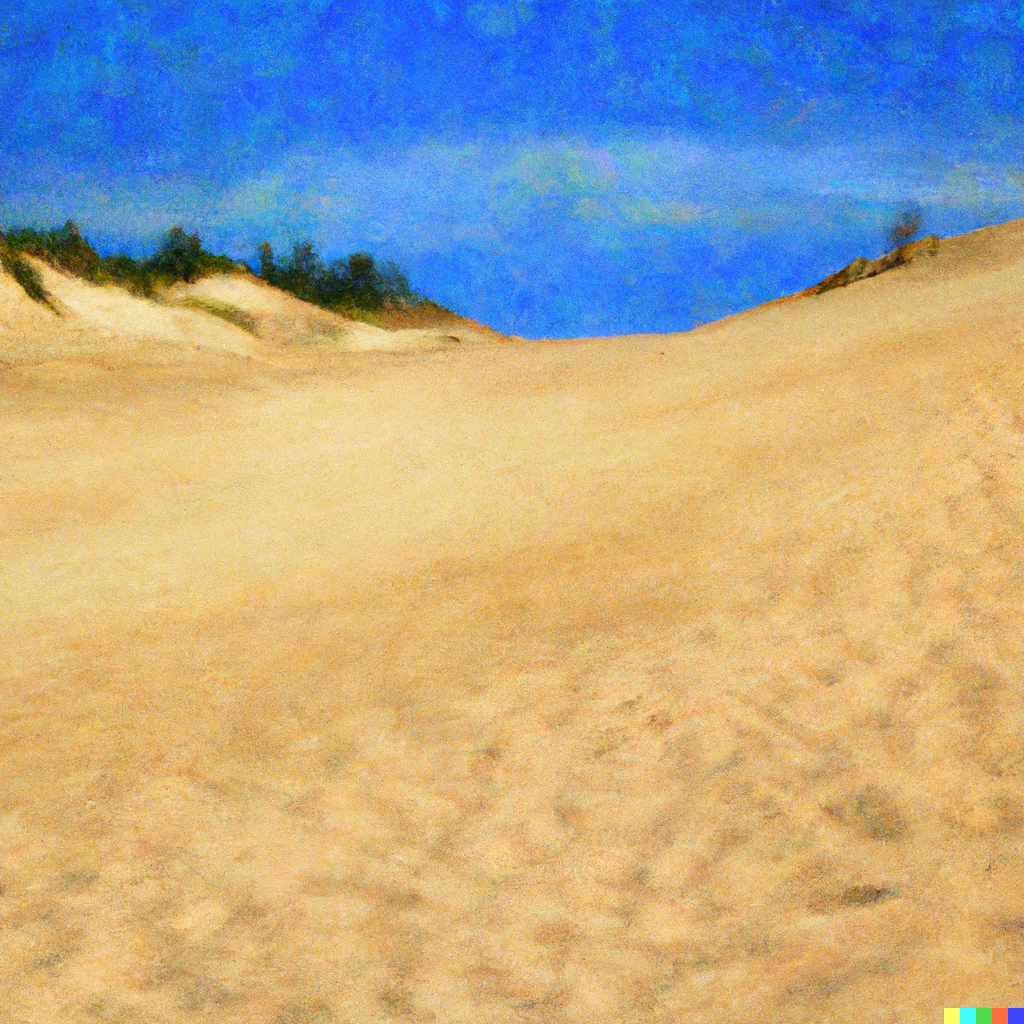} & \includegraphics[width=\linewidth]{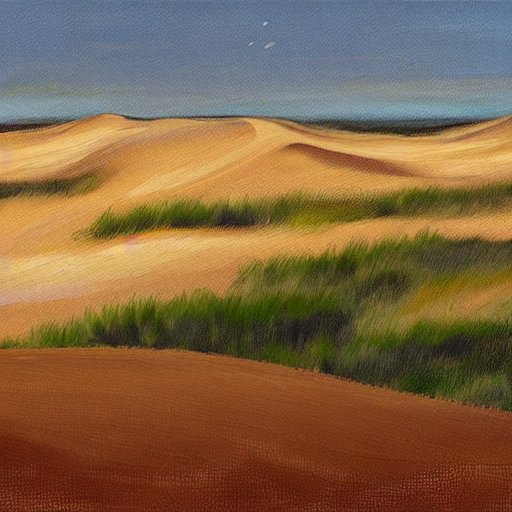} & \includegraphics[width=\linewidth]{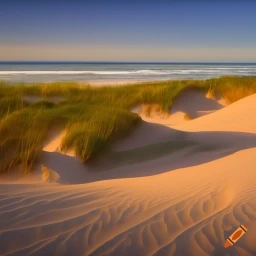} \\
    \bottomrule
    \end{tabular}%
\end{table}

\subsection{Results for Nuclear Power Prompts -- Promising Performance}
\label{sec:promising nuclear results}

As indicated in Section \ref{sec:Results for General Prompts}, the generative AI models have successfully generated accurate images in response to prompts related to the natural environment. Next, we examine the extent to which models are trained with prompts related to nuclear energy. We provide nuclear-related prompts to the models and analyze the outcomes to understand their proficiency in generating images in this specific domain.

In this exploration, we tested the 3 generative AI tools against four prompts, the results are shown in Table \ref{tab:promisingnucresults}. In our first prompt, we asked all 3 tools to produce an image of a ``Person who works in the nuclear industry.'' DALL-E 2 produced an image of a male with a mask working at a nuclear power plant, standing next to a single cooling tower. The image  appears very detailed and realistic, though showing only a cooling tower and not a reactor building. Additionally, the image does not accurately depict the attire of nuclear plant workers. DreamStudio produced two male workers in work attire and hard hats inside a nuclear power plant. Craiyon created a male in a hard hat, in front of an electrical grid. It did not directly produce anything related to a nuclear power plant, but did display a power transformer. Interestingly, each model only depicted men as nuclear plant workers, thus reproducing existing gender imbalances. It is also notable that DALL-E 2 and DreamStudio generated images of workers who appear to be Caucasian, whereas Craiyon generated an image of an ethnically ambiguous worker.

The second prompt we tested was ``Impact of Uranium mining on Indigenous Peoples’ traditional lands.'' DALL-E 2 produced an image of dry desert land with a small pond of water nearby, with cut-down trees. This image does not appear to be a Uranium mine, but is a high-quality image. DreamStudio produced a more accurate image of a Uranium mine, depicting rock and dirt excavated at different levels. It also showed animals and tools at the bottom of the image, inferring that these are Indigenous tools. Craiyon produced a technically accurate image of a Uranium mining, depicting different mining levels in a desert environment. Craiyon produced an image that is more of a drawing/painting, and not an image. However, Craiyon generated nothing related to ``Indigenous people". We further improvised this prompt by specifying Navajo instead of indigenous people. Therefore, the prompt was changed to "Impact of Uranium mining on Navajo traditional lands"; in this case, Craiyon and Dreamstudio could capture the landscape of Navajoland, indicating improvement in Craiyon performance as the prompt got more specific. Dreamstudio could also include a Uranium mine in the image. However, DALL-E produced an image of dry land, failing to generate both Uranium mine and Navajoland. 

The fourth prompt we tested was ``Wildlife near a nuclear plant.'' DALL-E 2 produced two ducks on the dirt around grass, with two cooling towers in the background. The detail of the ducks and the cooling towers are accurate, and looks close to reality. StableDiffusion generated a deer next to a cooling tower in long grass; the image looks noisy and grainy. Some features of StableDiffusion's generated image are not detailed, as the sky is not a palette of blues, there are no clouds or other background scenery, and the grass proportionally tall and one dimensional compared to the deer and cooling tower. However, this image still accurately represents the prompt. Next, Craiyon accurately produced two cooling towers; however, it attempted to generate an animal at the top of the smoke clouds. It is also worth noting that the steam exiting each cooling tower is in opposite directions; this is barely possible for steam to be carried in opposite directions by the wind. Despite this error, it was included in successful attempts, as it still accurately portrayed nuclear cooling towers and attempted to create an animal.

It seems that general nuclear energy prompts produce promising results; however, it was observed that nuclear prompts tend to almost always produce cooling towers. This could be due to the data sets used to train these generative models in that there is am availability of large number of cooling tower images on internet in comparison to images of other technical components. This suggests that all the explored models associate nuclear energy with cooling towers; however, it also suggests that these generative AI models do not have a thorough understanding of other components of nuclear power and nuclear power plants in areas outside of cooling towers.

\begin{figure}[!h]
    \centering
    \includegraphics[width=1.0\textwidth]{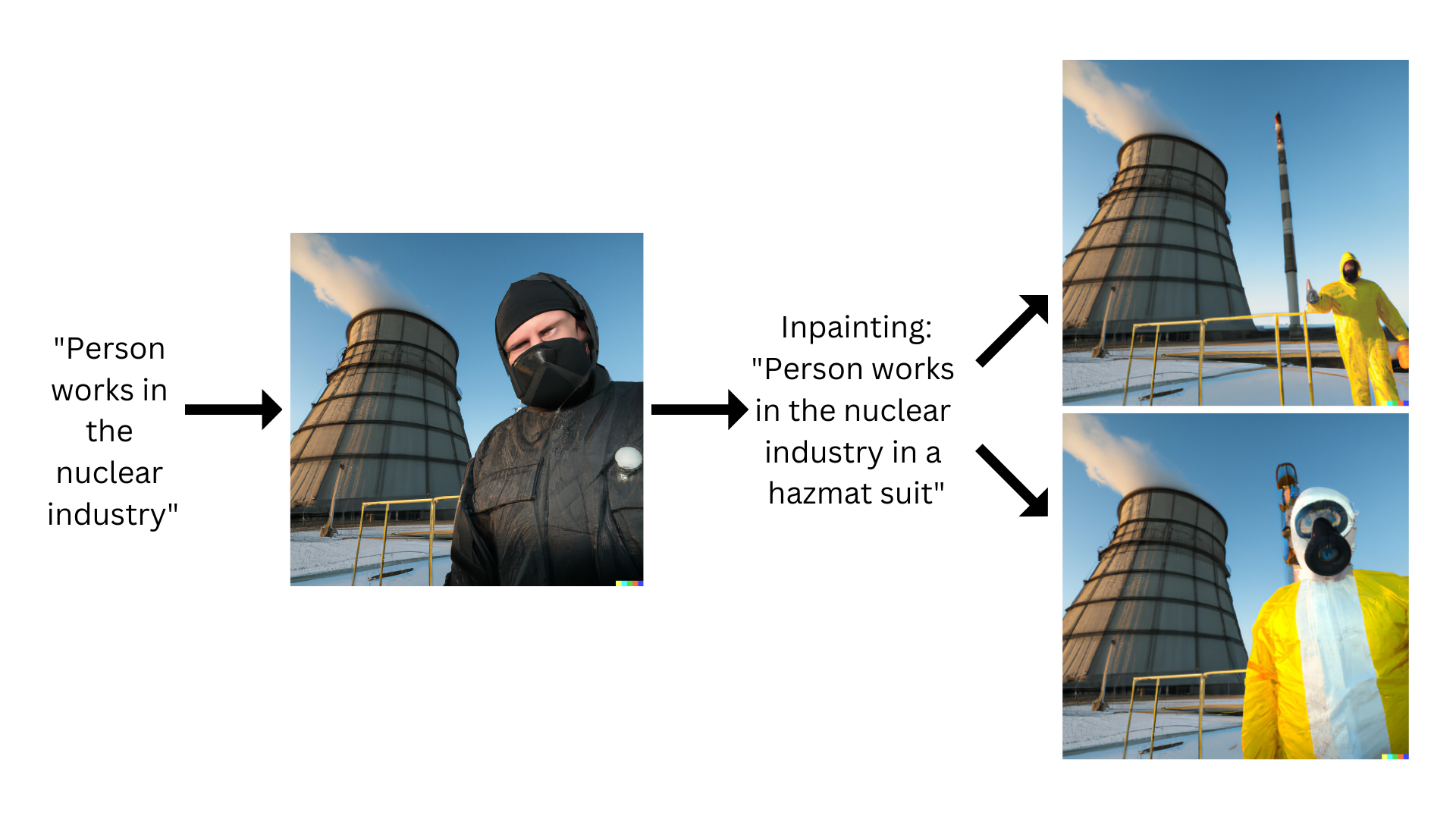}
    \caption{DALL-E 2 Inpainting for a nuclear energy prompt} 
    \label{fig:in}
\end{figure}

\begin{figure}[!h]
    \centering
    \includegraphics[width=\textwidth]{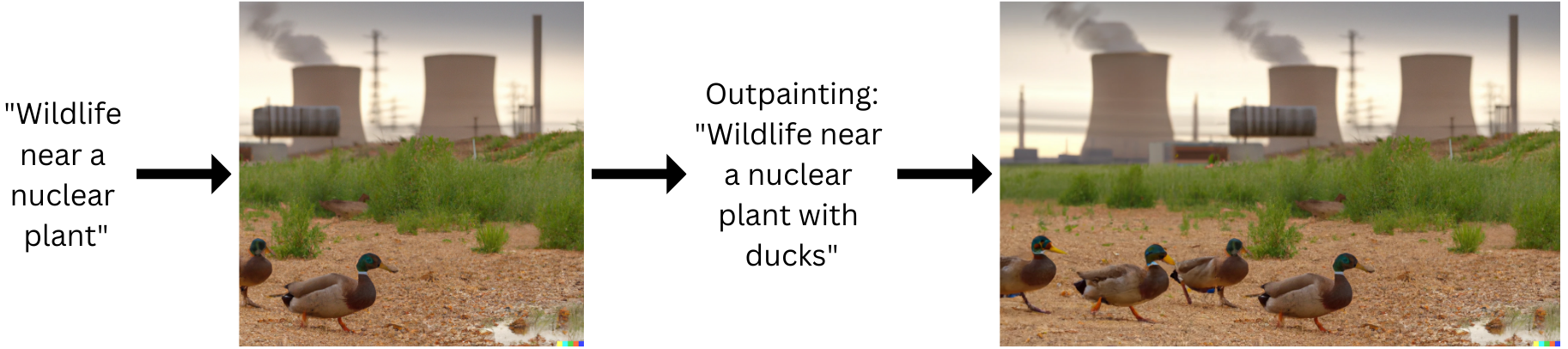}
    \caption{DALL-E 2 Outpainting for a nuclear energy prompt}
    \label{fig:out}
\end{figure}

\begin{table}[!h]
\caption{Promising results from nuclear related prompts}
\label{tab:promisingnucresults}%
\centering
    \begin{tabular}{m{2.7cm}|m{4.2cm}|m{4.2cm}|m{4.2cm}}
    \toprule
    Prompts & DALL-E 2 & DreamStudio & Craiyon \\
    \midrule
     Person works in the nuclear industry & \includegraphics[width=\linewidth]{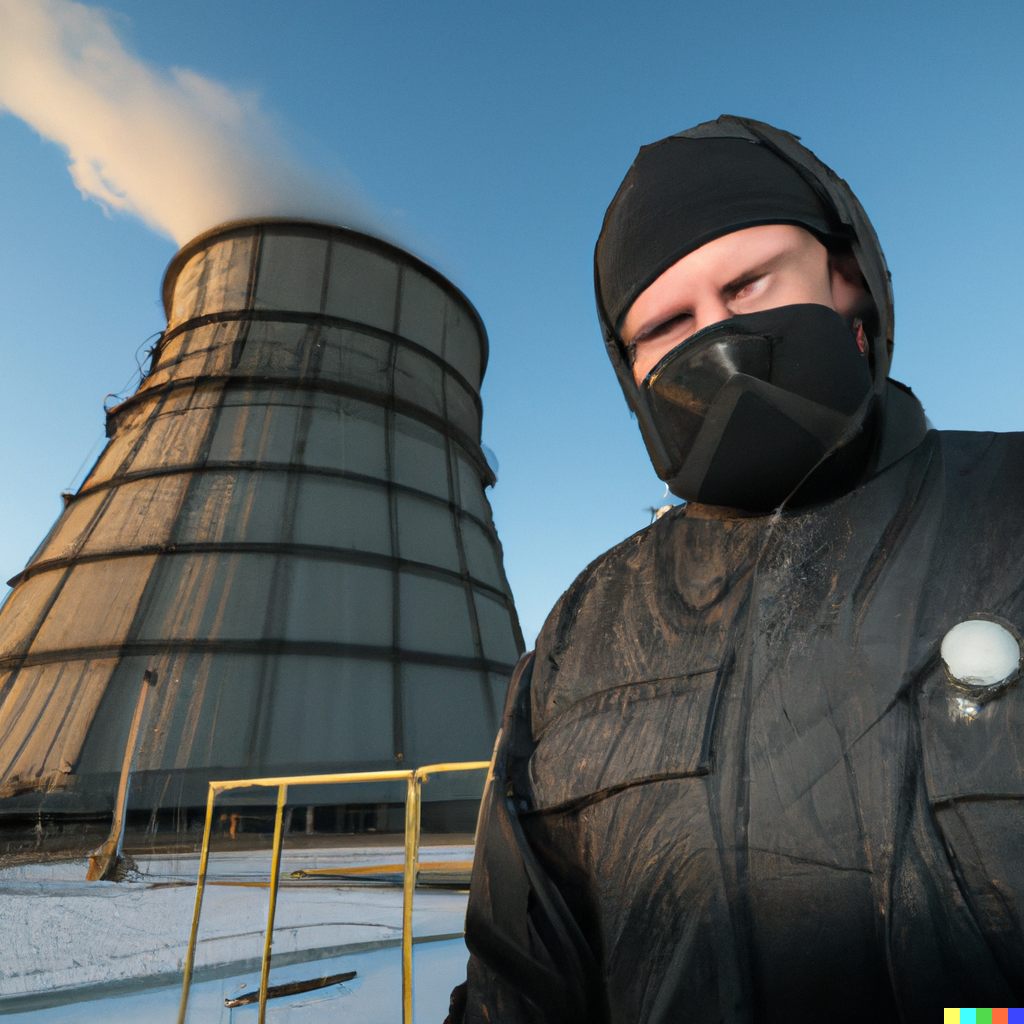} & \includegraphics[width=\linewidth]{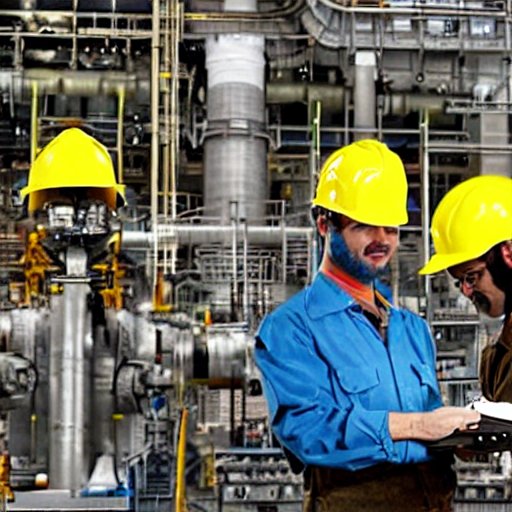} & \includegraphics[width=\linewidth]{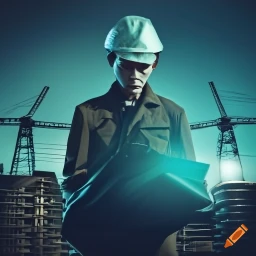} \\
    \midrule
    Impact of Uranium mining on Indigenous Peoples' traditional lands & \includegraphics[width=\linewidth]{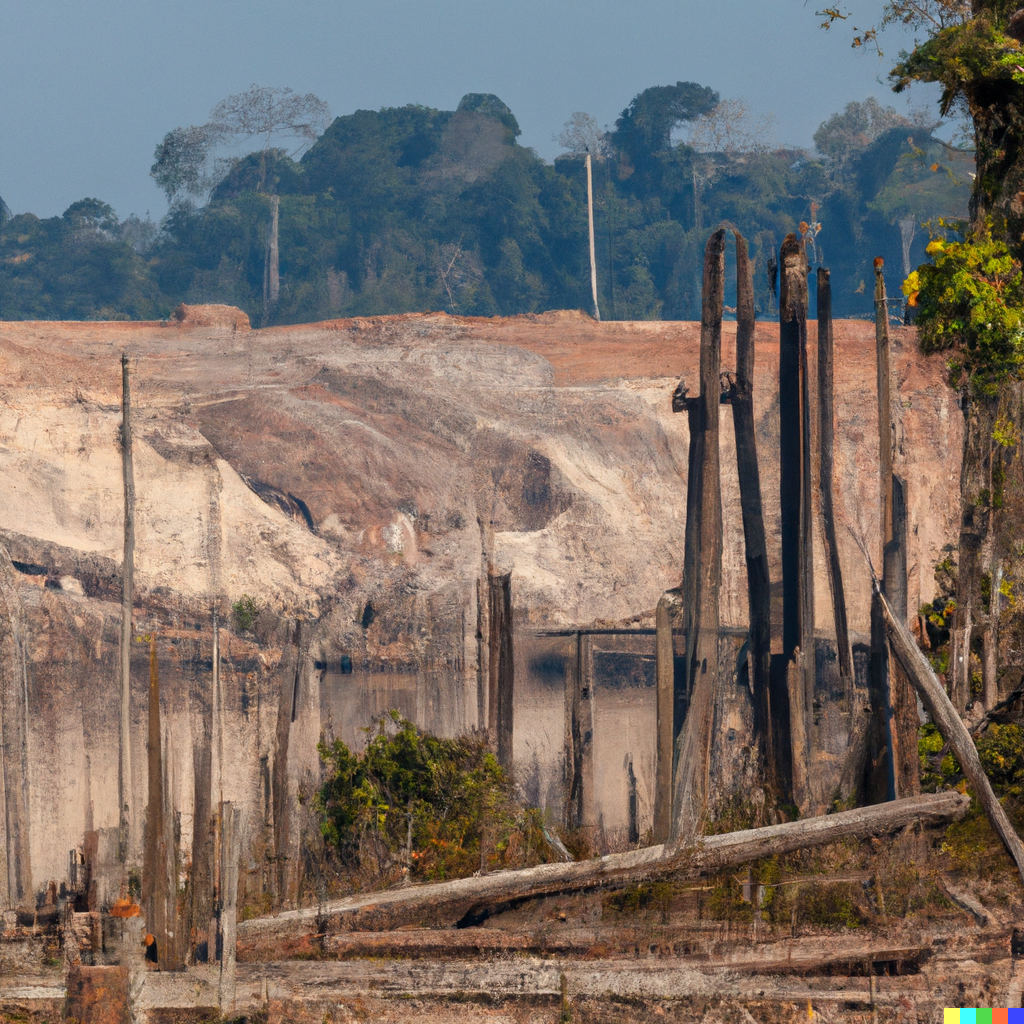} & \includegraphics[width=\linewidth]{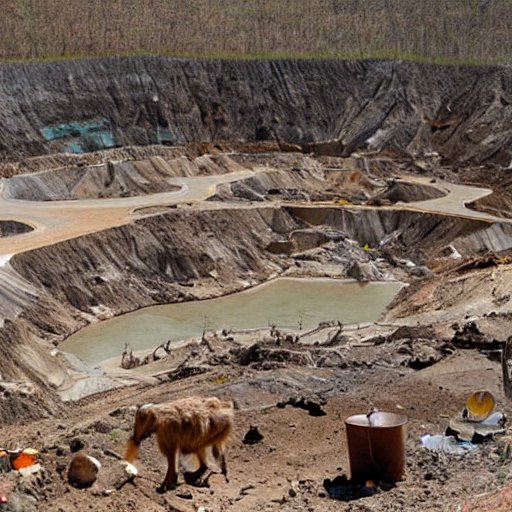} & \includegraphics[width=\linewidth]{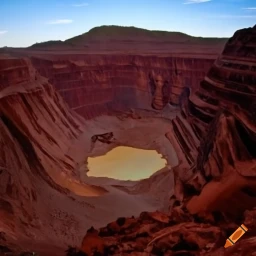} \\
    \midrule
    Impact of Uranium mining on Navajo traditional lands & \includegraphics[width=\linewidth]{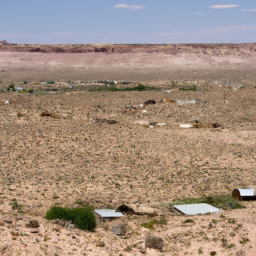} & \includegraphics[width=\linewidth]{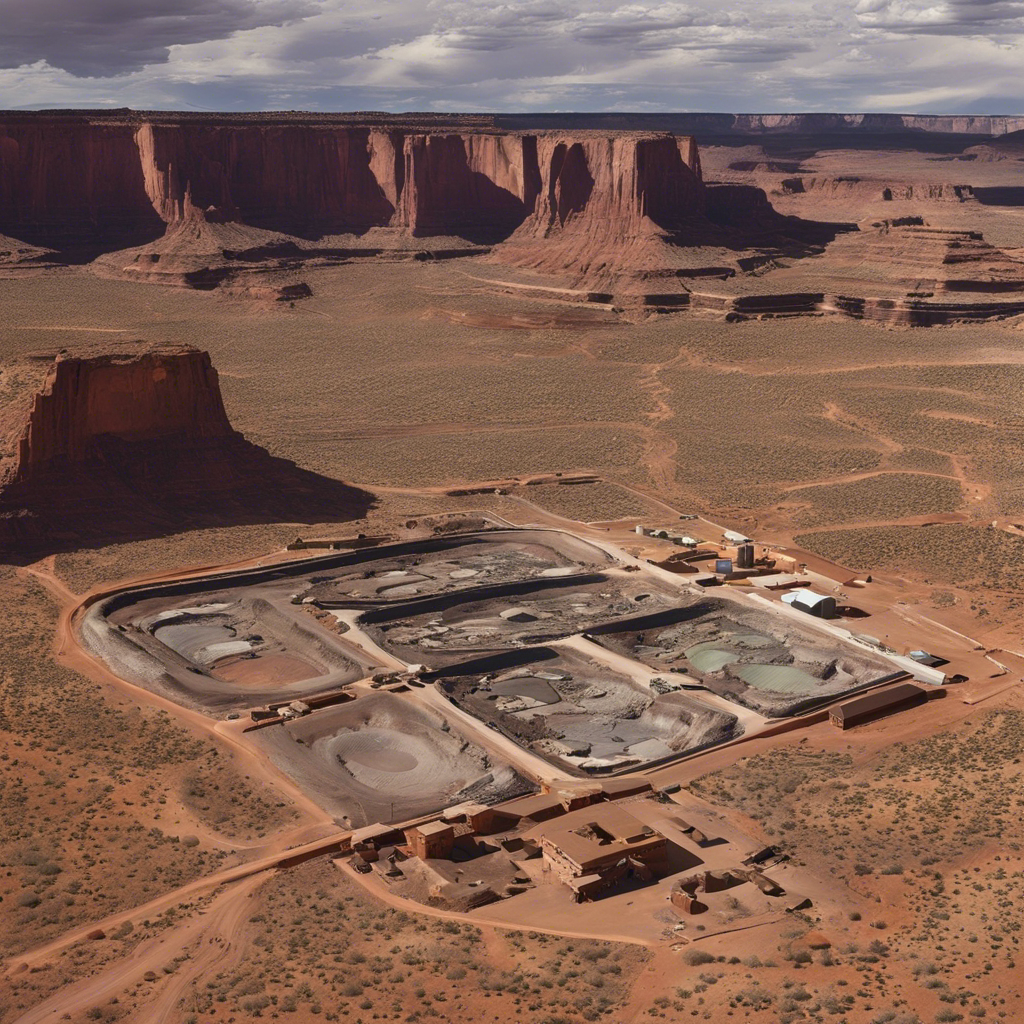} & \includegraphics[width=\linewidth]{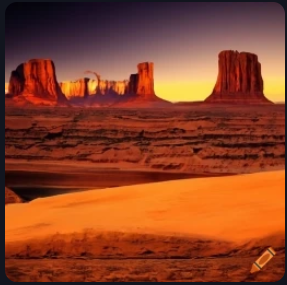}
     \\
    \midrule
    Wildlife near a nuclear plant & \includegraphics[width=\linewidth]{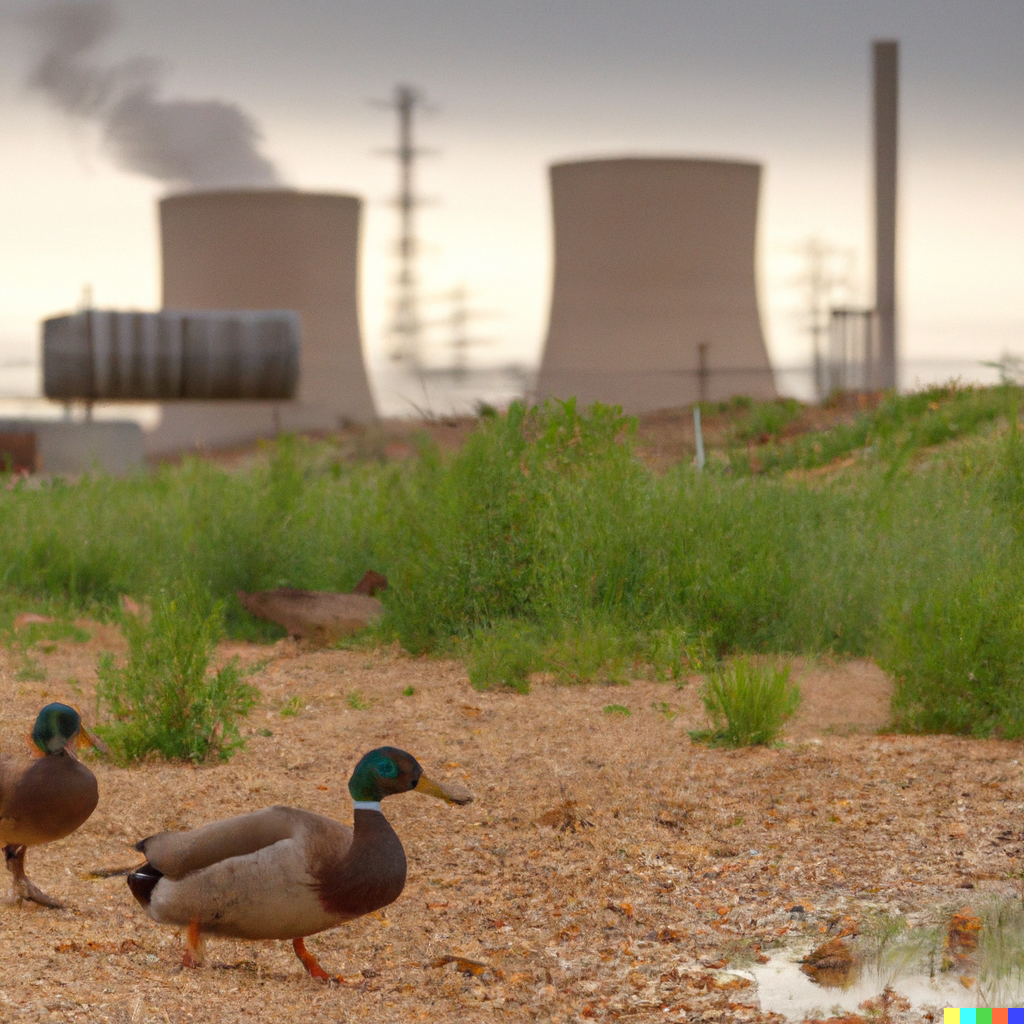} & \includegraphics[width=\linewidth]{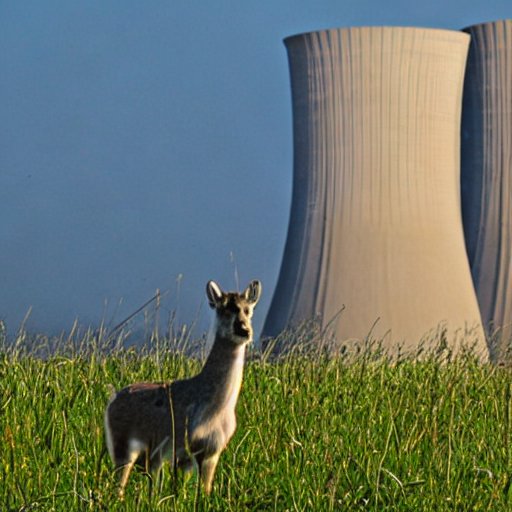} & \includegraphics[width=\linewidth]{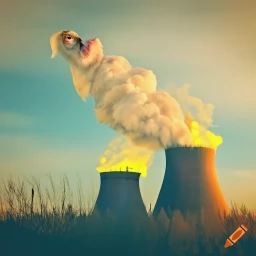}
 \\

    \bottomrule
    
    \end{tabular}%
\end{table}
In the next phase of this work, we went beyond image generation and explored image editing capabilities using inpainting and outpainting functionalities. With the image that DALL-E 2 generated to the prompt ``Person works in the nuclear industry,'' we used the inpainting prompt ``Person Near a nuclear power plant in a hazmat suit.'' The resulting images produced a man in a hazmat suit in Figure \ref{fig:in}.\\
Next, we used outpainting feature with the DALL-E 2 image from the prompt "Wildlife near a nuclear plant." This was used to expand the borders of the image on the left side of the image. The outpainting algorithm added additional ducks, while adding a third cooling tower. The resulting images are shown in Figure \ref{fig:out}. 

\begin{table}[!h]
\caption{Poor Nuclear Results.}
\label{tab:poor nuclear}%
\centering
    \begin{tabular}{m{2.7cm}|m{4.2cm}|m{4.2cm}|m{4.2cm}}
    \toprule
    Prompts & DALL-E 2 & DreamStudio & Craiyon \\
    \midrule
     China and Nuclear & \includegraphics[width=\linewidth]{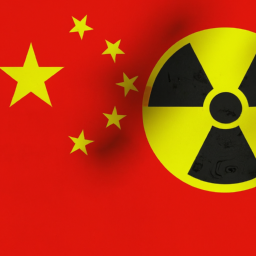} & \includegraphics[width=\linewidth]{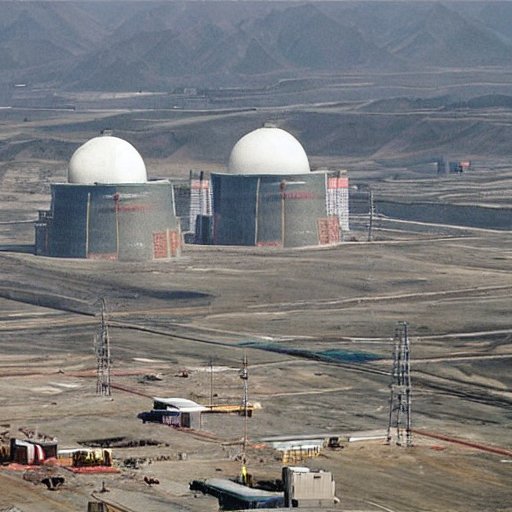} & \includegraphics[width=\linewidth]{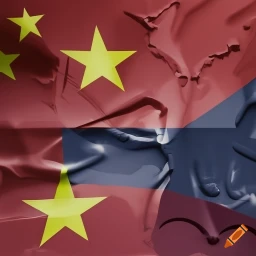} \\
    \midrule
     Display radioactive nuclear waste & \includegraphics[width=\linewidth]{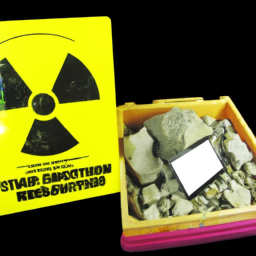} & \includegraphics[width=\linewidth]{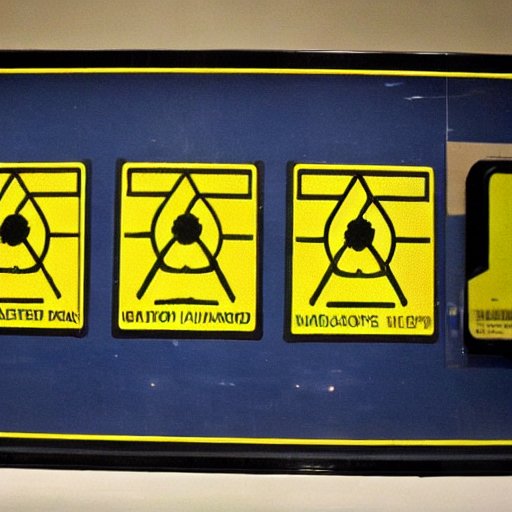} & \includegraphics[width=\linewidth]{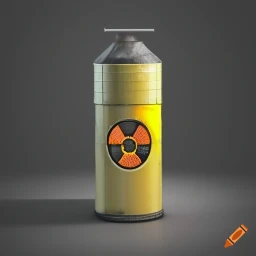} \\
    \midrule
    Create a functional diagram of a nuclear reactor core & \includegraphics[width=\linewidth]{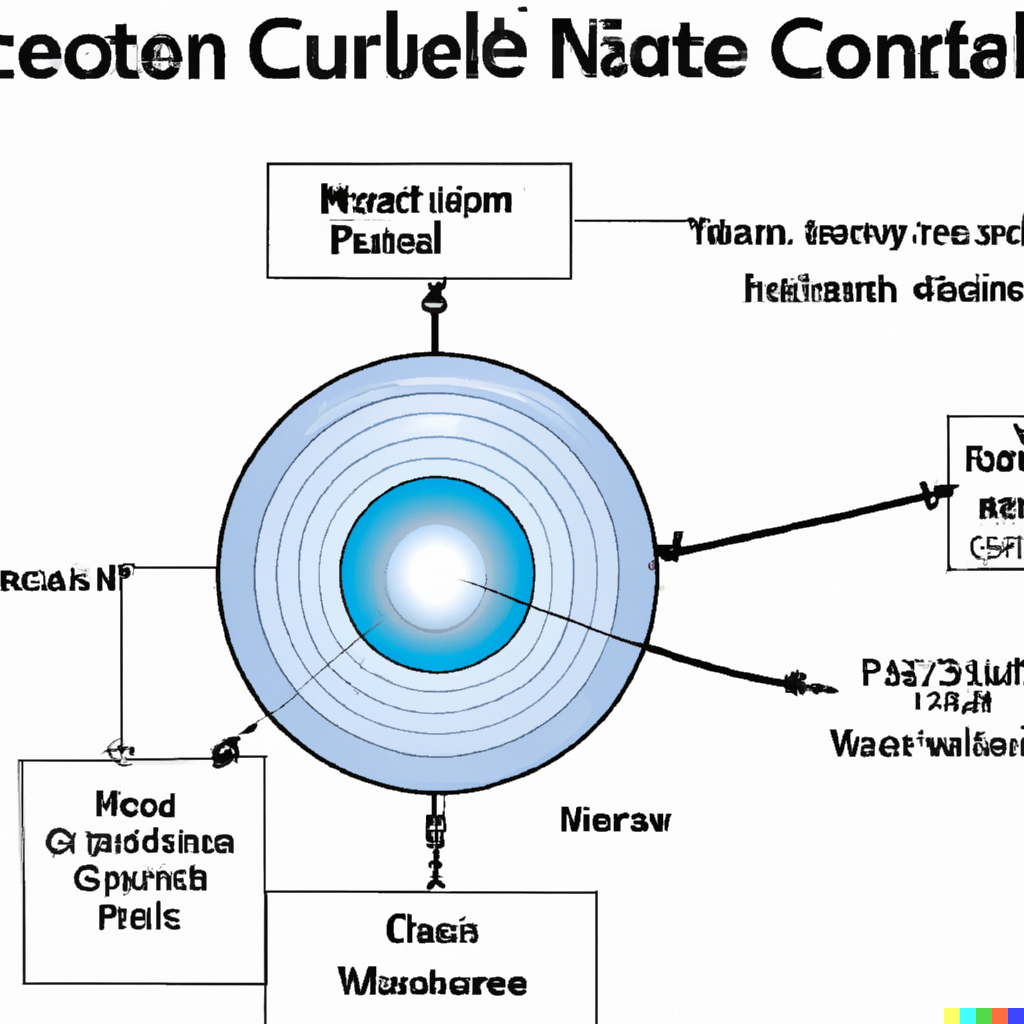} & \includegraphics[width=\linewidth]{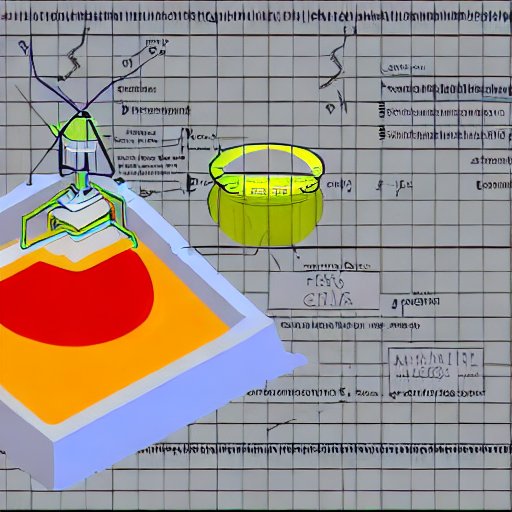} & \includegraphics[width=\linewidth]{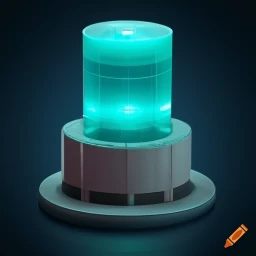} \\
    \bottomrule
    \end{tabular}%
\end{table}

\subsection{Results for Nuclear Power Prompts -- Poor Performance}
In Section \ref{sec:promising nuclear results}, we examined successful cases of generative AI nuclear energy applications. However, models also occasionally generated poor images depending on the prompts as shown in Table \ref{tab:poor nuclear}. The first unsuccessful prompt was ``China and nuclear.'' DALL-E 2 produced a flag similar in color and pattern to the Chinese flag, and included the atomic nuclear symbol on the flag. DreamStudio produced 2 extremely wide cooling towers, but there is nothing indicative of China in this picture. Craiyon produced another flag similar to the Chinese flag, but has an unusual blue stripe. There is nothing indicative about nuclear in this image. Text to image generative AI models struggled to link countries to nuclear. 

Next, we tried the prompt ``Display radioactive waste.'' DALL-E 2 produced an image of a crate with what our researchers believed to be stones inside the box, with an atomic logo and incomprehensible text on a lid. DreamStudio has boxes with a pattern and text on a yellow background; however, this image did not appear to be related to the prompt. Craiyon has the closest depiction of nuclear waste (even though still very far), of an atomic logo being in a cylindrical container. None of these images are a correct depiction of nuclear waste. 

Our third prompt was ''Create a functional diagram of a nuclear reactor core.'' DALL-E 2 showed a nuclear reactor core from the top down and got the circle shape right. This image had text that is not English, and appears meaningless. The diagram is also not technically accurate. DreamStudio attempted to create a diagram of a reactor core; the words are not legible and the diagram is difficult to see; this is also not correct on a technological level. Craiyon did not create a diagram, and it created a blue light cylinder on a grey base. Overall, none of these images show a correct diagram of a nuclear reactor core. 

\begin{figure}[!h]
    \centering
    \includegraphics[width=\textwidth]
    {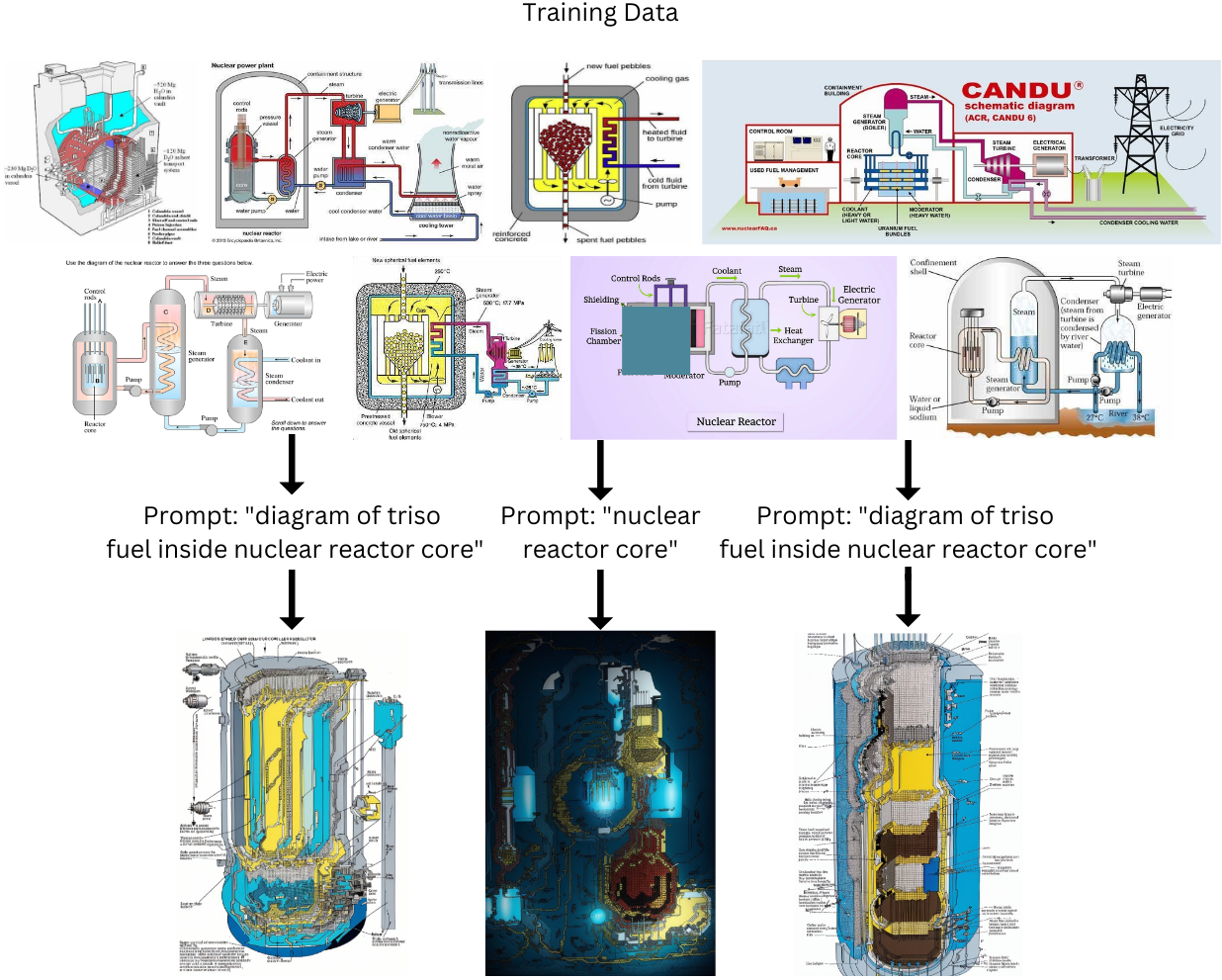}
    \caption{Training set and output of various prompts to provide a sketch of a nuclear reactor core} 
    \label{fig:trainingdata}
\end{figure}

While general nuclear prompts produced promising results, anything technical or requiring words produced meaningless results. In a failed attempt to create better results, we used Leonardo.AI's model training with a small data set of 8 images to make a more accurate nuclear diagram.  Figure \ref{fig:trainingdata} illustrates the training set as well as the diagrams produced. The diagrams generated by the trained model appeared plausible at first glance; however, upon closer examination numerous issues surfaced. Firstly, the characters displayed on the diagram remained intricate gibberish, and the nuclear fuel rods that should have been situated within the reactor core were absent. Though images containing such technical and specialized content prioritize precise information transmission over creativity, none of those tools satisfied the criteria. It seems that more extensive training and meticulous adjustments are necessary.

\subsection{Results for Prompt Engineering}

To test the efficacy of prompt engineering as indicated in Figure \ref{fig:pealgo}, a set of 7 prompts related to nuclear engineering were collected from a set of subjects. Table \ref{tab:dalle2pe}, Table \ref{tab:craiyonpe}, and Table \ref{tab:dspe} display the prompt engineering results by DALLE-2, Craiyon, and Dreamstudio models, respectively. Each table contains the original prompt provided by the analyst, the image associated with the original prompt, the modified prompt by prompt engineering, and the modified image associated with that modified prompt.

In the case of DALL-E, Table \ref{tab:dalle2pe} illustrates a combination of promising and unsatisfactory outcomes following prompt engineering. Notably, prompts 1, 2, and 4 related to the control room, spent fuel pool, and fission reaction exhibited considerable improvement, while the others remained inaccurate. Prompt 1, for instance, resulted in a modified image of the control room that closely resembled an actual nuclear control room, however it omitted the nuclear reactor core. Despite prompt 2 omitting nuclear waste and spent fuel, the modified version still portrayed a realistic image of the spent fuel pool, where nuclear waste is temporarily stored after being discharged from the reactor for cooling. Prompt 4, focusing on the fission reaction, displayed atoms splitting into smaller atoms, reflecting the process of fission. However, prompt 4's results still contained unreadable wording and gibberish language. Prompts 3, 5, 6, and 7 remained considerably distant from reality. For instance, prompt 6 failed to depict the nuclear fuel pellet and nuclear fuel rod in the context of a birthday event.

Tables \ref{tab:craiyonpe} and \ref{tab:dspe} depict the outcomes of Crayion and DreamStudio, revealing inferior performance compared to DALL-E across both original and modified prompts. The generated images from these models consistently exhibit unrealistic characteristics. Notably, the only instances of improvement are observed in prompt 1 for Crayion and DreamStudio, associated with the nuclear control room. Additionally, prompt 2 for DreamStudio yields somewhat realistic images of the spent fuel pool, with both the original and modified versions displaying a fair quality.

\begin{small}
\begin{longtable} {|m{0.5cm}|m{2.5cm}|m{3.5cm}|m{2.5cm}|m{3.5cm}|}
\caption{DALLE-2 Results with Prompt Engineering}
\label{tab:dalle2pe}\\
\toprule
No.& Prompts & Original Image & Modified Prompts & Modified Image \\
\hline
1 & Nuclear reactor control room with reactor core visible from a glass, people are sitting in front of the controls. & \includegraphics[width=3.5cm, height=3.5cm]{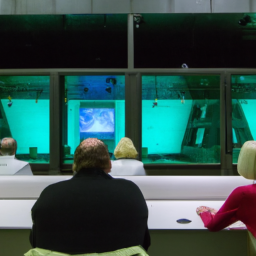} & Nuclear reactor control room with scientists working in it. The room contains multiple gauges and levers. & \includegraphics[width=3.5cm, height=3.5cm]{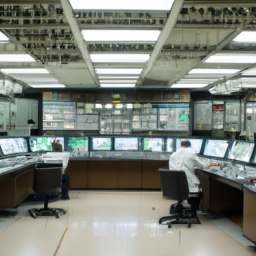} \\
\hline
2 & Spent nuclear fuel in a cooling pool. & \includegraphics[width=3.5cm, height=3.5cm]{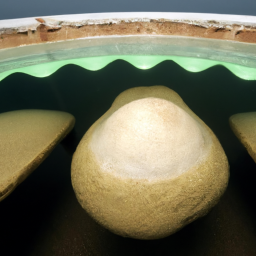} & Display cooling pool. It is a deep empty pool inside the nuclear power plant. & \includegraphics[width=3.5cm, height=3.5cm]{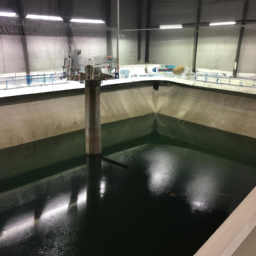} \\
\hline
3 & A beam of neutrons hitting a slab of iron. & \includegraphics[width=3.5cm, height=3.5cm]{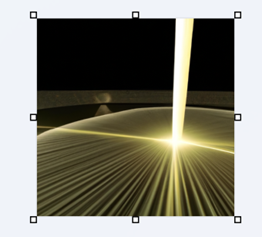} & Diagram of beam of neutrons represented as blue rays falling from top on slab of iron represented as black rectangle. & \includegraphics[width=3.5cm, height=3.5cm]{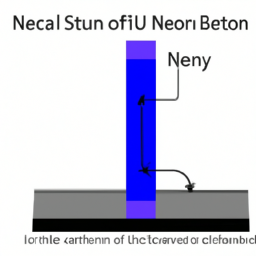} \\
\hline
4 & Fission reaction in the nuclear reactor core. & \includegraphics[width=3.5cm, height=3.5cm]{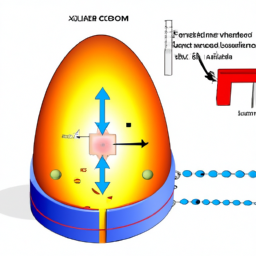} & Show an atom splitting into two smaller atoms., Display three atoms, one atom is bigger than other two atoms. & \includegraphics[width=3.5cm, height=3.5cm]{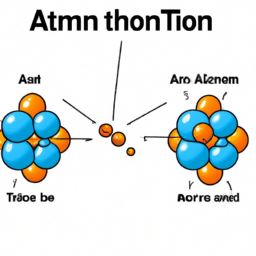}\\
\hline
5 & Show the radiation shielding of a nuclear reactor. & \includegraphics[width=3.5cm, height=3.5cm]{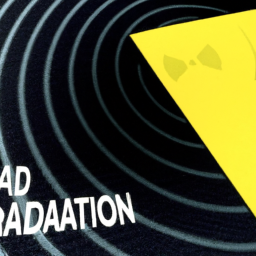} & Display technical diagram of a dome covering nuclear reactor. & \includegraphics[width=3.5cm, height=3.5cm]{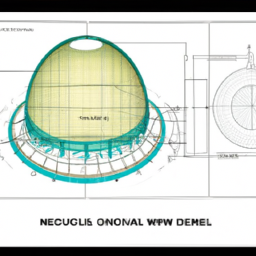}\\
\hline
6 & 10 pressurized water reactor fuel pellets are celebrating the birthday of a new nuclear fuel rod in a water pool.& \includegraphics[width=3.5cm, height=3.5cm]{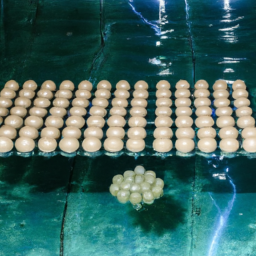} & Display 10 water reactor fuel pellet. It looks like a black cylinder. Display a fuel rod, it looks like a steel rod. Put cake and birthday cap in background. & \includegraphics[width=3.5cm, height=3.5cm]{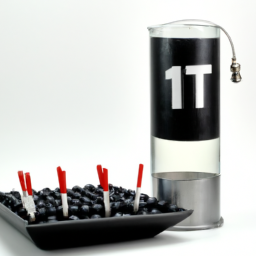}\\
\hline
7 & Primary side of a PWR including reactor, pressurizer, reactor coolant pump, and steam generator & \includegraphics[width=3.5cm, height=3.5cm]{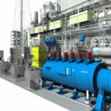} & A image of the reactor coolant system and the primary side of a pressurized water reactor, which could include components like the reactor core, pressurizer, steam generator, and coolant pumps. & \includegraphics[width=3.5cm, height=3.5cm]{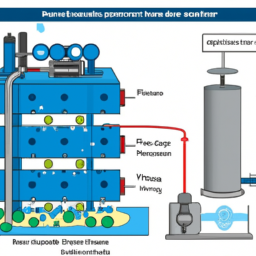}\\
\bottomrule
\end{longtable}

\begin{longtable} {|m{0.5cm}|m{2.5cm}|m{3.5cm}|m{2.5cm}|m{3.5cm}|}
\caption{Craiyon Results with Prompt Engineering}
\label{tab:craiyonpe} \\
\toprule
No.& Prompts & Original Image & Modified Prompts & Modified Image \\
\hline
1 & Nuclear reactor control room with reactor core visible from a glass, people are sitting in front of the controls. & \includegraphics[width=3.5cm, height=3.5cm]{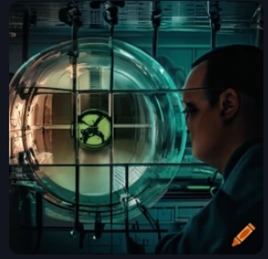} & technically accurate Nuclear reactor control room with scientists working in it The room contains multiple gauges and levers Lifelike & \includegraphics[width=3.5cm, height=3.5cm]{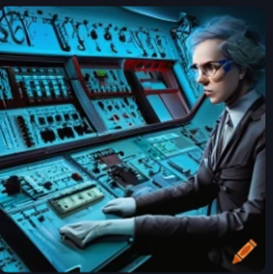} \\
\hline
2 & Spent nuclear fuel in a cooling pool. & \includegraphics[width=3.5cm, height=3.5cm]{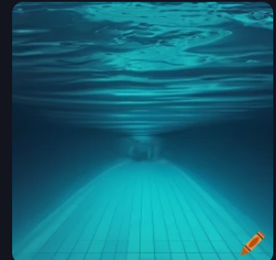} & Finely detailed photograph Display cooling pool It is a deep empty pool inside the nuclear power plant. & \includegraphics[width=3.5cm, height=3.5cm]{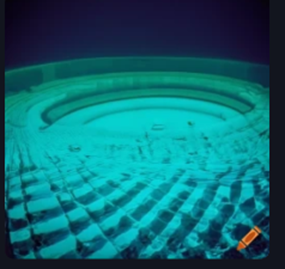} \\
\hline
3 & A beam of neutrons hitting a slab of iron. & \includegraphics[width=3.5cm, height=3.5cm]{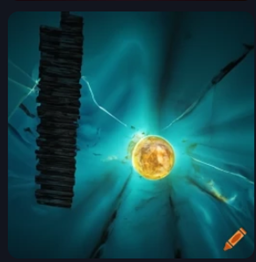} & Carefully executed photograph Diagram of beam of neutrons represented as blue rays falling from top on slab of iron represented as black rectangle. Lifelike & \includegraphics[width=3.5cm, height=3.5cm]{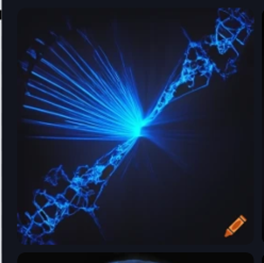} \\
\hline
4 & Fission reaction in the nuclear reactor core. & \includegraphics[width=3.5cm, height=3.5cm]{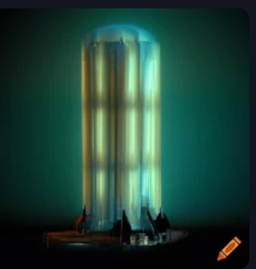} & Show an atom splitting into two smaller atoms., Display three atoms, one atom is bigger than other two atoms. & \includegraphics[width=3.5cm, height=3.5cm]{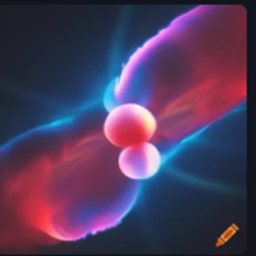}\\
\hline
5 & Show the radiation shielding of a nuclear reactor. & \includegraphics[width=3.5cm, height=3.5cm]{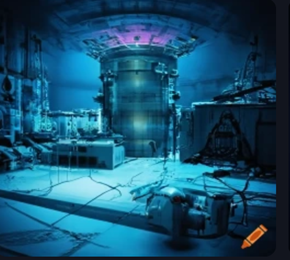} & highly detailed Display technical diagram of a dome covering nuclear reactor. Accurate representation & \includegraphics[width=3.5cm, height=3.5cm]{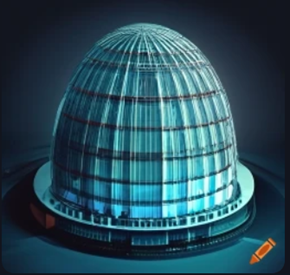}\\
\hline
6 & 10 pressurized water reactor fuel pellets are celebrating the birthday of a new nuclear fuel rod in a water pool.& \includegraphics[width=3.5cm, height=3.5cm]{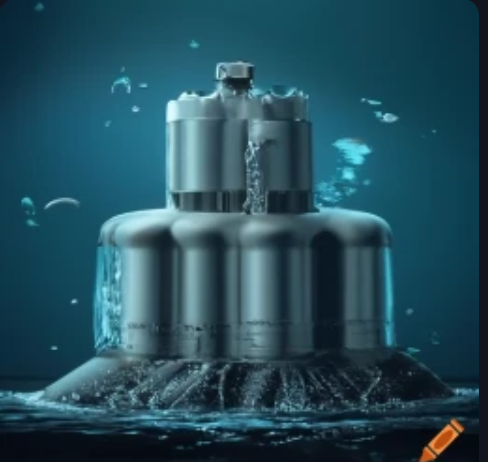} & highly detailed Display 10 water reactor fuel pellet. It looks like a black cylinder. Display a fuel rod, it looks like a steel rod. Put cake and birthday cap in background. Accurate representation & \includegraphics[width=3.5cm, height=3.5cm]{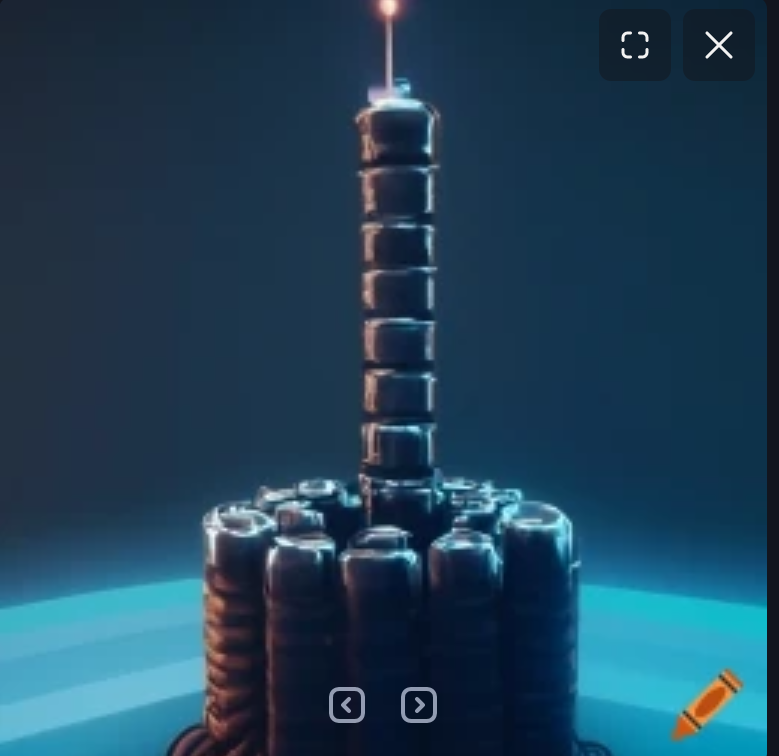}\\
\hline
7 & Primary side of a PWR including reactor, pressurizer, reactor coolant pump, and steam generator & \includegraphics[width=3.5cm, height=3.5cm]{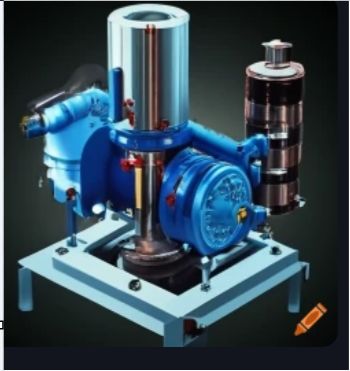} & A detailed and realistic image of the reactor coolant system and the primary side of a pressurized water reactor, which could include components like the reactor core, pressurizer, steam generator, and coolant pumps. &   \includegraphics[width=3.5cm, height=3.5cm]{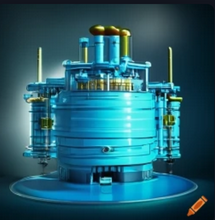}\\
\bottomrule
\end{longtable}

\begin{longtable} {|m{0.5cm}|m{2.5cm}|m{3.5cm}|m{2.5cm}|m{3.5cm}|}
\caption{DreamStudio Results with Prompt Engineering}
\label{tab:dspe}\\
\toprule
No. & Prompts & Original Image & Modified Prompts & Modified Image \\
\hline
1 & Nuclear reactor control room with reactor core visible from a glass, people are sitting in front of the controls. & \includegraphics[width=3.5cm, height=3.5cm]{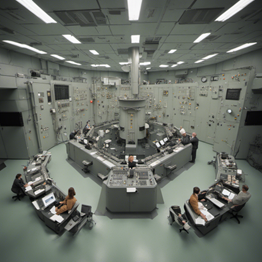} & Nuclear reactor control room with scientists working in it. The room contains multiple gauges and levers. & \includegraphics[width=3.5cm, height=3.5cm]{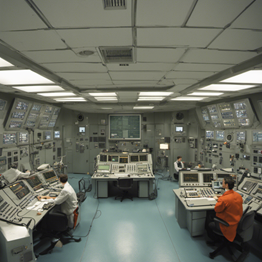} \\
\hline
2 & Spent nuclear fuel in a cooling pool. & \includegraphics[width=3.5cm, height=3.5cm]{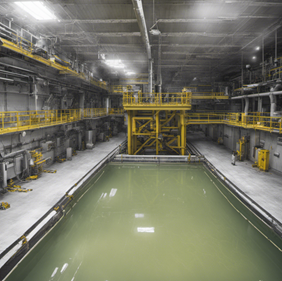} & Display cooling pool. It is a deep empty pool inside the nuclear power plant. & \includegraphics[width=3.5cm, height=3.5cm]{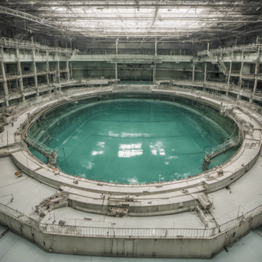} \\
\hline
3 & A beam of neutrons hitting a slab of iron. & \includegraphics[width=3.5cm, height=3.5cm]{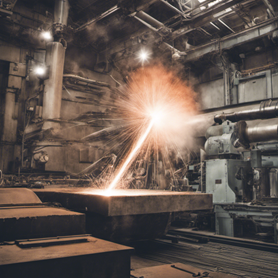} & Diagram of beam of neutrons represented as blue rays falling from top on slab of iron represented as black rectangle. & \includegraphics[width=3.5cm, height=3.5cm]{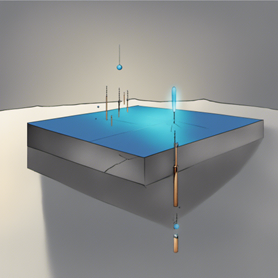} \\
\hline
4 & Fission reaction in the nuclear reactor core. & \includegraphics[width=3.5cm, height=3.5cm]{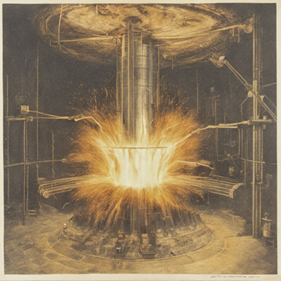} & Show an atom splitting into two smaller atoms. & \includegraphics[width=3.5cm, height=3.5cm]{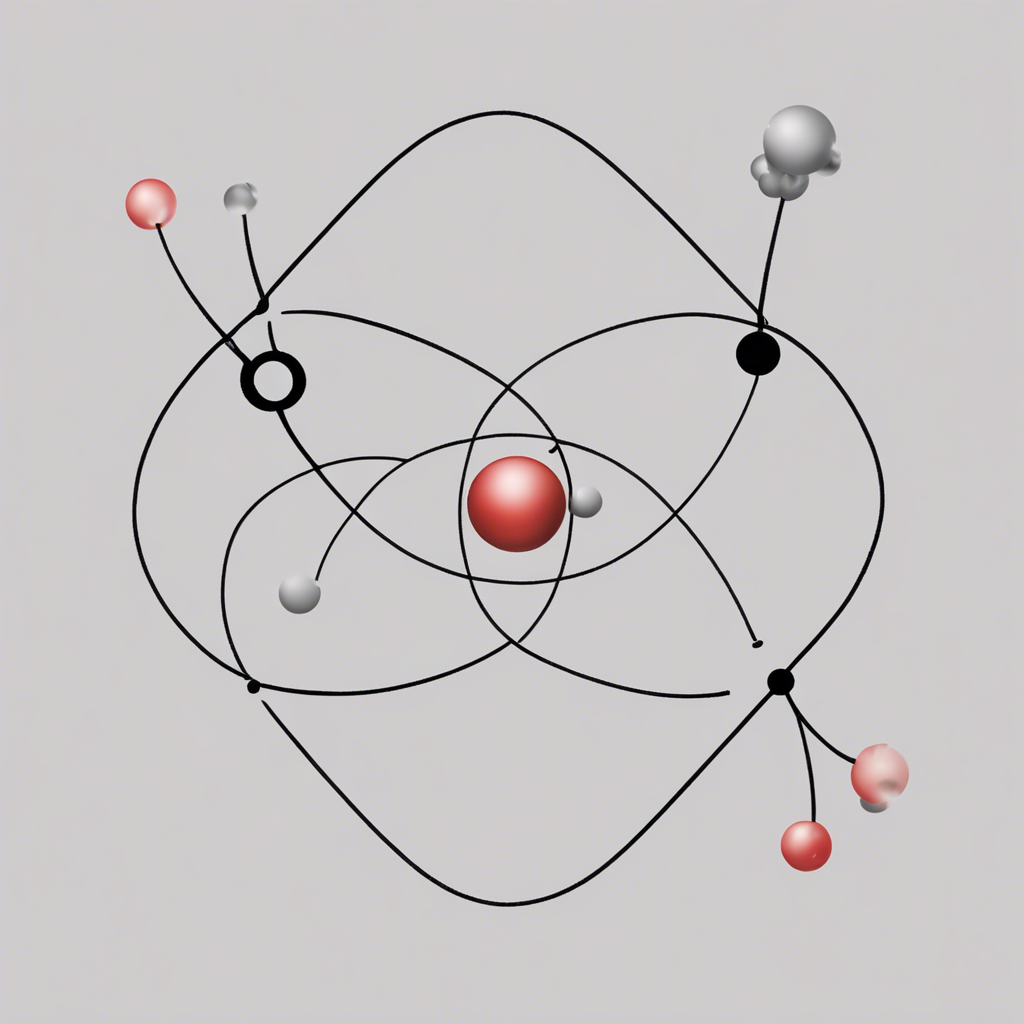}\\
\hline
5 & Show the radiation shielding of a nuclear reactor. & \includegraphics[width=3.5cm, height=3.5cm]{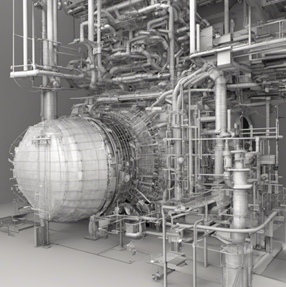} & Display technical diagram of a dome covering nuclear reactor. & \includegraphics[width=3.5cm, height=3.5cm]{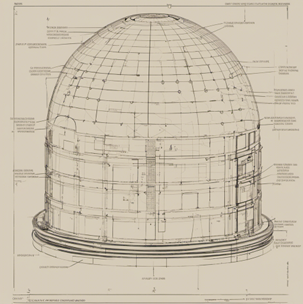}\\
\hline
6 & 10 pressurized water reactor fuel pellets are celebrating the birthday of a new nuclear fuel rod in a water pool.& \includegraphics[width=3.5cm, height=3.5cm]{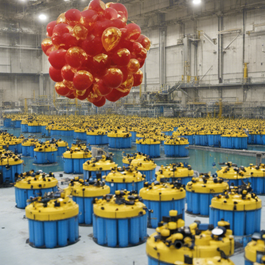} & Display 10 water reactor fuel pellet. It looks like a black cylinder. Display a fuel rod, it looks like a steel rod. Put cake and birthday cap in background. & \includegraphics[width=3.5cm, height=3.5cm]{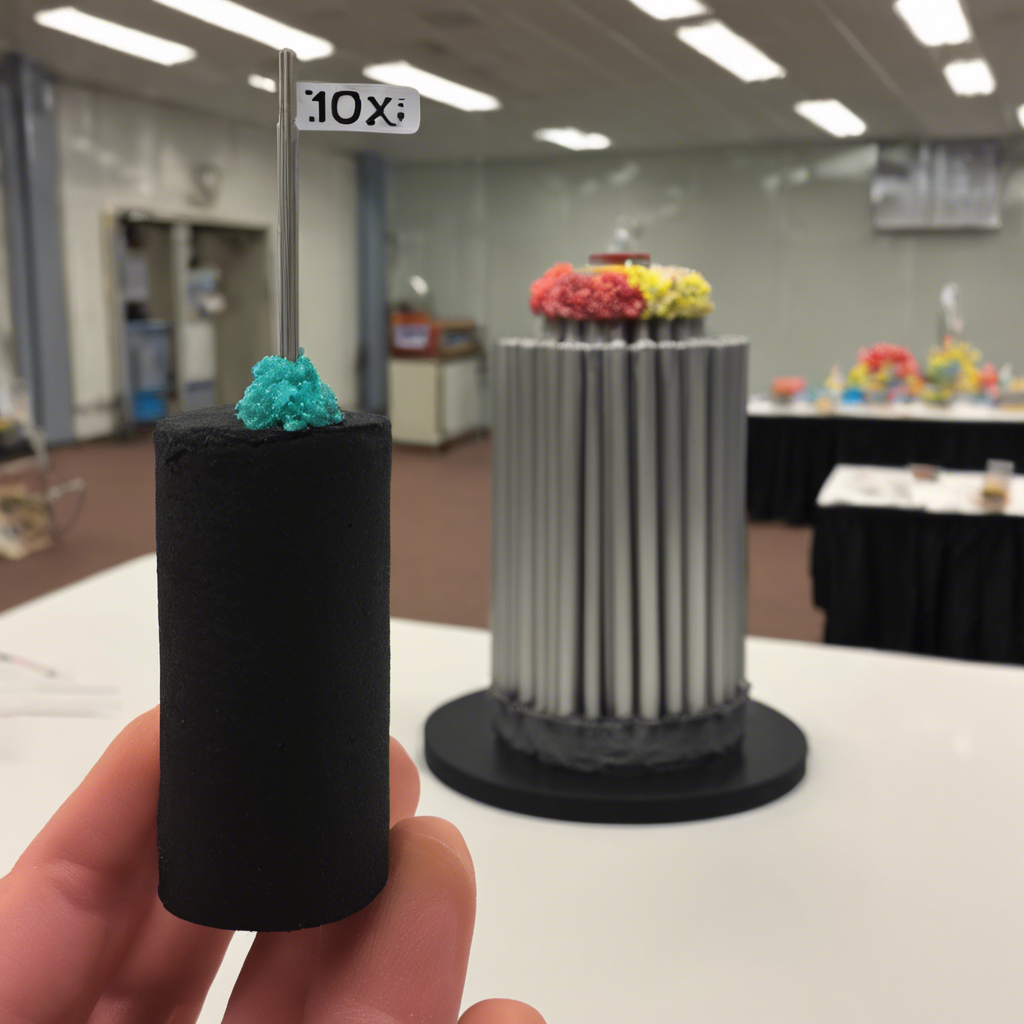}\\
\hline
7 & Primary side of a PWR including reactor, pressurizer, reactor coolant pump, and steam generator & \includegraphics[width=3.5cm, height=3.5cm]{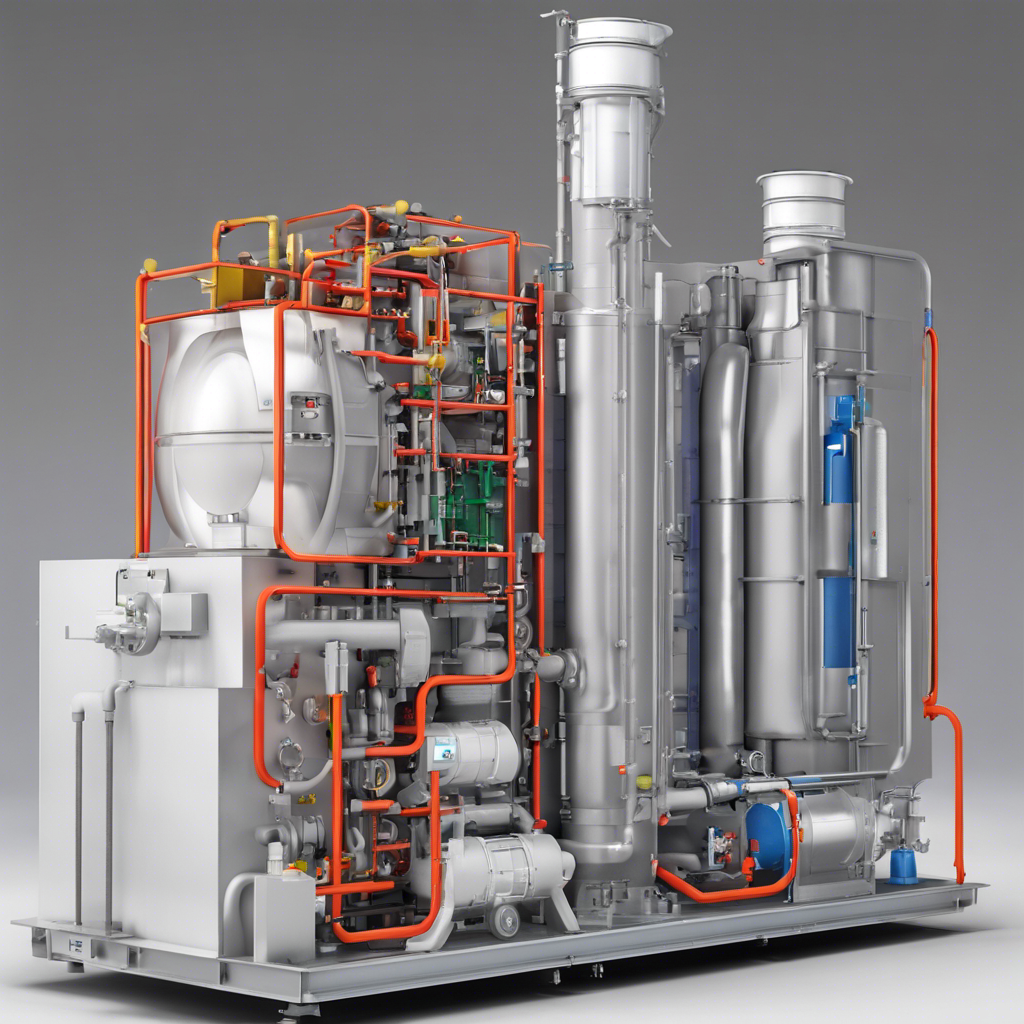} & A image of the reactor coolant system and the primary side of a pressurized water reactor, which could include components like the reactor core, pressurizer, steam generator, and coolant pumps. & \includegraphics[width=3.5cm, height=3.5cm]{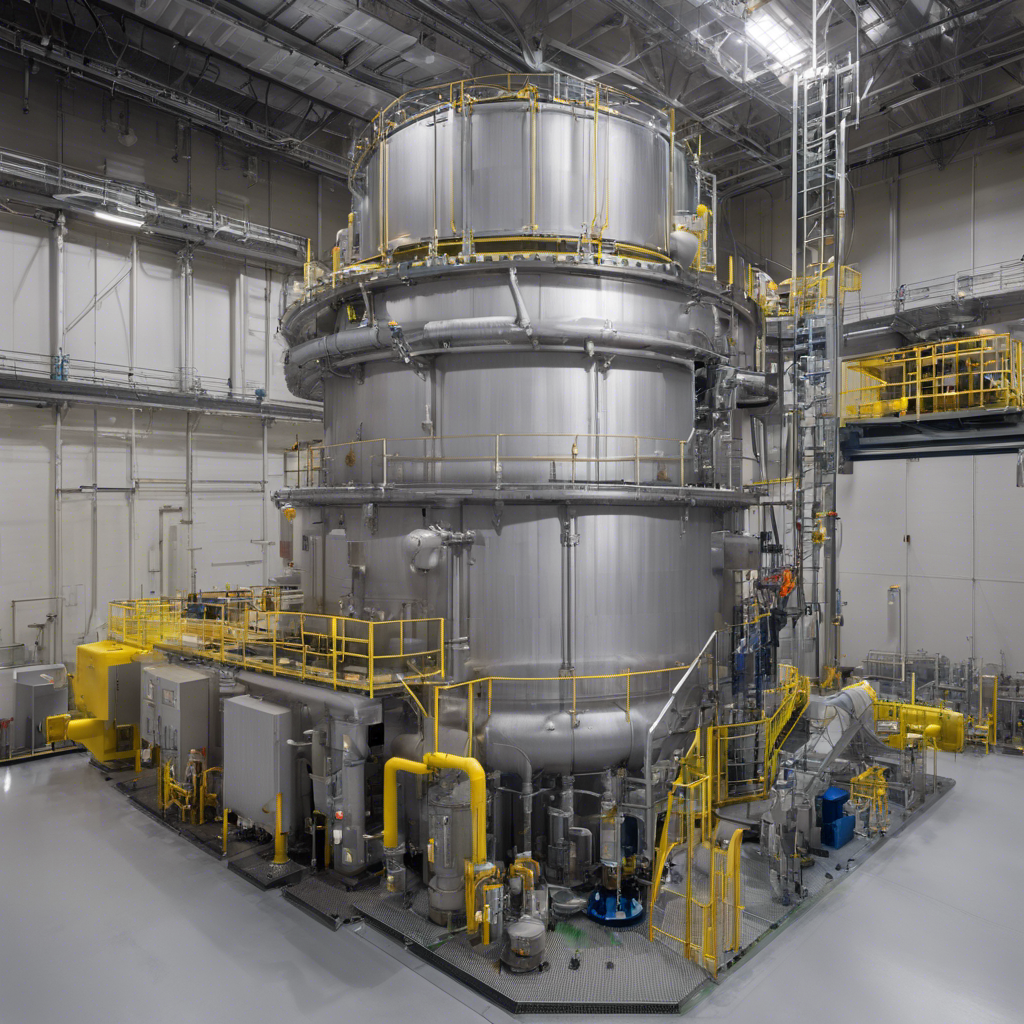}  \\
\bottomrule
\end{longtable}
\end{small}

\subsection{Discussion}
The main goal of our research focuses on creating realistic and accurate images that could accurately depict nuclear energy in both technical and non-technical ways to reflect the reliability of generative AI models in a scientific context. Text-to-image systems appear to successfully depict only the cooling towers in a nuclear plant; they struggle with technical details related to nuclear power plants. For example, as shown in Figure \ref{fig:trainingdata}, all models could not produce a diagram of a nuclear reactor, even with some training data templates that depict real nuclear reactor cores. Additionally, we noticed that radioactive waste was portrayed incorrectly by DALL-E 2 and DreamStudio, and only Craiyon depicted a barrel, which is still far from how the radioactive wastes and their storage casks look like. Such outcomes indicate that models have not yet been adequately trained on data related to nuclear power. Such a conclusion might be also true for other scientific disciplines. Consequently, the development of a generative AI specialized in nuclear necessitates the acquisition of a greater volume of nuclear-energy-related data. Ensuring sufficient high-quality training data must undoubtedly be incorporated into future work. 


Comparing all three models, DALLE-2 gave the best results with prompt engineering. It was also noticed that DALL-E 2 generated better images when only a few number of subjects are present in the prompt, otherwise, different objects interpolated into each other. For instance, in case of Prompt 2 at Table \ref{tab:dalle2pe}, optimal results are obtained by removing nuclear fuel from the original prompt and instead describing about cooling pool in detail. Further, in case of prompt 1, optimal results could be obtained by removing nuclear reactor core from the prompt. This pattern was observed in all the three models. From the prompt engineering results, it can be observed that generative AI models give better results for the nuclear components that have a substantial number of images present on the internet, such as depicting cooling towers or nuclear reactor control rooms than the prompts referring to subjects which have comparatively less images (e.g., steam generator, reactor core, fuel rod). Further, the results can be improved by giving visual cues to the model in layman's language, as in the case of cooling fuel in prompt 2 of Table \ref{tab:dalle2pe}. However, all generative AI models are still not able to comprehend the technical terms, relying on the appearance description provided in most of the cases.

Additionally, it is important to note that all images presented in this paper are the initial images generated through generative AI. We have made this decision in light of concerns about cherry-picking results and potentially biasing performance by selecting the best images after multiple attempts. That said, we conducted a single execution for each model, and from the approximately 3 to 4 images obtained from that run, we selected the images that we deemed to be of the highest quality to include in this paper. It should be noted that among our multidisciplinary team of nuclear engineers, AI, and data scientists, the researchers who chose the prompt and verified its quality had a nuclear engineering background.

Through this study, we have also identified several common issues that models encounter during image generation. All generative AI tools struggled with accurately drawing human faces. This may be due to the numerous facial expressions and facial variations humans have, which would result in having an extremely large database of human faces in order to accurately portray the human face. As Table \ref{tab:promisingnucresults}'s prompt 1 shows, one can recognize that they are humans, but the faces are off. The man's right eye is malformed. For prompt 3 on the same table, eyes were completely overshadowed by the hat. The generative AI tools also perpetuate prevailing biases related to gender and employment within the nuclear energy sector and inadequately depict indigenous environments, which have traditionally served as locations for resource extraction and the disposal of nuclear waste by energy industries.

Furthermore, words were presented as nonsense. Despite entering an English prompt, the characters presented in the generated image were intricate symbols rather than alphabetic letters. There could be multiple reasons for this phenomenon, yet the most plausible explanation is the insufficient training of the model in depicting textual content directly as images. In other words, while the machine has acquired the capability to illustrate the entity referred to by the text ``nuclear power reactor," it has not been trained to produce the exact image representation of the text ``nuclear power reactor" itself.

In summary, after this study's exploration of various generative AI models with a specific focus towards nuclear engineering in both a technical and non-technical sense,  we have found that a nuclear-specific generative AI model is needed as current models lack the technical expertise to accurately illustrate nuclear topics besides the stereotypes. Beyond that, we have found that for the generative AI models studied they struggle with producing images with readable words, and human faces despite slight improvements after applying prompt engineering. However, we should emphasize that the concerns we found here are specific to the models we tested even though the pool we tested is still large.

\FloatBarrier 
\section{Conclusions}
\label{sec:conc}

In the context of humanity, who must concurrently address energy crisis and climate change, communication between the general public and experts regarding green energy has become more crucial than ever. 
As stated by Veera et al. \cite{vimpari2023adapt}, one of the current slogans in the field of artificial intelligence is `Adapt or Die.' In this regard, energy experts should now harness generative AI to create synergistic effects in their communication with the public.
In this study, we explored various generative AI models in search for ones that accurately depict scientific and nuclear energy prompts from both a technical and non-technical perspective.
Among 20 tools, we narrowed our focus to DALL-E 2, Craiyon, and DreamStudio for their promising results on general nuclear prompts. Through our exploration, we found that all the models we studied struggle with creating images of technical nuclear objects such as ``nuclear reactor core.'' Specifically, we found that the models struggle with complex objects and technical terminologies in general. We noticed an overabundance of nuclear cooling towers during the research. While cooling towers are the most noticeable for the general public when it comes to nuclear energy, it does not accurately portray nuclear energy, which further suggests that a nuclear energy-specific generative AI is needed. This could also be true for other energy systems (e.g., renewable). 

Prompt Engineering techniques were applied to further optimize the prompt and generate desired images. It was noticed that improved results can be obtained by giving highly specified prompts to the generative model, along with substantial descriptions regarding the visual appearance of the prompt subject. However, improvement was mostly seen in result of prompts having a large number of related images present on the internet or containing common more general terms, like deer grazing near a cooling tower. The model still was unable to comprehend the technical terms related to nuclear engineering or generate images when multiple nuclear objects are present in the prompt. Though these models are not satisfactory as of now, they may be significantly improved if they are trained on a large data set of nuclear-related images. Furthermore, when accompanied by efforts to optimize prompts, the model performance is likely to improve even further. 

In light of these findings, our research team's future works are as follows.  It is evident that for a specialized text-to-image generative model for nuclear energy, a greater accumulation of pertinent training data is imperative. The variance in data volume across domains introduces substantial performance disparities. As demonstrated in our study, as shown in Table \ref{tab:generalresults}, all three tools generated images of near-perfect quality for the text ``Bunny.'' In contrast, as evident from Table \ref{tab:poor nuclear}, when it comes to nuclear expertise-related content, they produce perplexing images. Ultimately, increased exposure to certain texts during training allows for the refinement of image generation, indicating that the more exposure, the more accurate the imagery becomes. Consequently, our research team recognizes the necessity for nuclear-centered generative AI development and intends to pursue this as part of our future work. In addition, gender, race, and ethnicity inclusive set of images would reduce the bias these tools carry. Such a specialized tool will then be tested through social experiments with the public to obtain realistic prompts regarding public concerns about nuclear power and clean energy policy. 

\section*{Acknowledgment}
Our team is grateful to the Fastest Path to Zero at the University of Michigan for sponsoring this work under project number (G028455). Furthermore, we express our gratitude to the graduate students and researchers affiliated with the Artificial Intelligence and Multiphysics Simulations (AIMS) group at the University of Michigan. Meredith Eaheart, Kamal Abdulraheem, Leo Tunkle, Umme Nabila, and Omer Erdem have suggested sample prompts that aided in the testing of the tools and methods employed in this study.

\section*{CRediT Author Statement}

\begin{itemize}
    \item \textbf{Veda Joynt}: Methodology, Software, Validation, Data Curation, Formal analysis, Visualization, Investigation, Writing - Original Draft. 
    \item \textbf{Jacob Cooper}:  Methodology, Software, Validation, Data Curation, Formal analysis, Visualization, Investigation, Writing - Original Draft. 
    \item \textbf{Naman Bhargava}: Methodology, Software, Validation, Data Curation, Formal analysis, Visualization, Investigation, Writing - Original Draft. 
    \item \textbf{Katie Vu}: Methodology, Software, Validation, Data Curation, Formal analysis, Visualization, Investigation, Writing - Original Draft. 
    \item \textbf{O Hwang Kwon}: Methodology, Software, Validation, Data Curation, Formal analysis, Visualization, Investigation, Writing - Original Draft. 
    \item \textbf{Todd R. Allen}: Conceptualization, Funding acquisition, Project administration, Writing - Review and Edit. 
    \item     \textbf{Aditi Verma}: Conceptualization, Methodology, Funding acquisition, Writing - Review and Edit. 
    \item \textbf{Majdi I. Radaideh}: Conceptualization, Methodology, Investigation, Funding acquisition, Supervision, Project administration, Writing - Review and Edit. 
\end{itemize}


\bibliographystyle{elsarticle-num}
\setlength{\bibsep}{0pt plus 0.3ex}

{
 \bibliography{references}}





\end{sloppypar}
\end{document}